\documentclass[12pt]{article}

\usepackage{nips15submit_e,times,amsmath,amssymb,graphicx}

\usepackage{graphicx}
\usepackage{algpseudocode}
\usepackage{algorithm}
\usepackage{comment}

\RequirePackage[numbers]{natbib}
\RequirePackage[colorlinks,citecolor=blue,urlcolor=blue]{hyperref}

\usepackage{cases}
\usepackage{subcaption}
\usepackage{tabularx}
\usepackage{url}

\newtheorem{theorem}{Theorem}

\newtheorem{corollary}[theorem]{Corollary}

\newtheorem{definition}{Definition}

\let\hat\widehat

\newcommand{\K}{\mathbb K}

\title{Statistical Inference Using Mean Shift Denoising}

\author{
Yunhua Xiang \\
Department of Statistics\\
University of Washington\\
\texttt{xiangyh@uw.edu} \\
\And
Yen-Chi Chen \\
Department of Statistics\\
University of Washington\\
\texttt{yenchic@uw.edu} \\
}

\nipsfinalcopy

\begin{document}
\maketitle

%

%

%
%

\begin{abstract}
In this paper, we study how the mean shift algorithm can be used to denoise a dataset. 
We introduce a new framework to analyze the mean shift algorithm as a denoising approach by viewing the algorithm as an operator on a distribution function. 
We investigate how the mean shift algorithm changes the distribution and show that data points shifted by the mean shift 
concentrate around high density regions of the underlying density function. 
By using the mean shift as a denoising method, we enhance the performance of several clustering techniques,
improve the power of two-sample tests,
and obtain a new method for anomaly detection.
\end{abstract}

\section{Introduction}

\begin{figure}
	\center
	\includegraphics[width=1.6in]{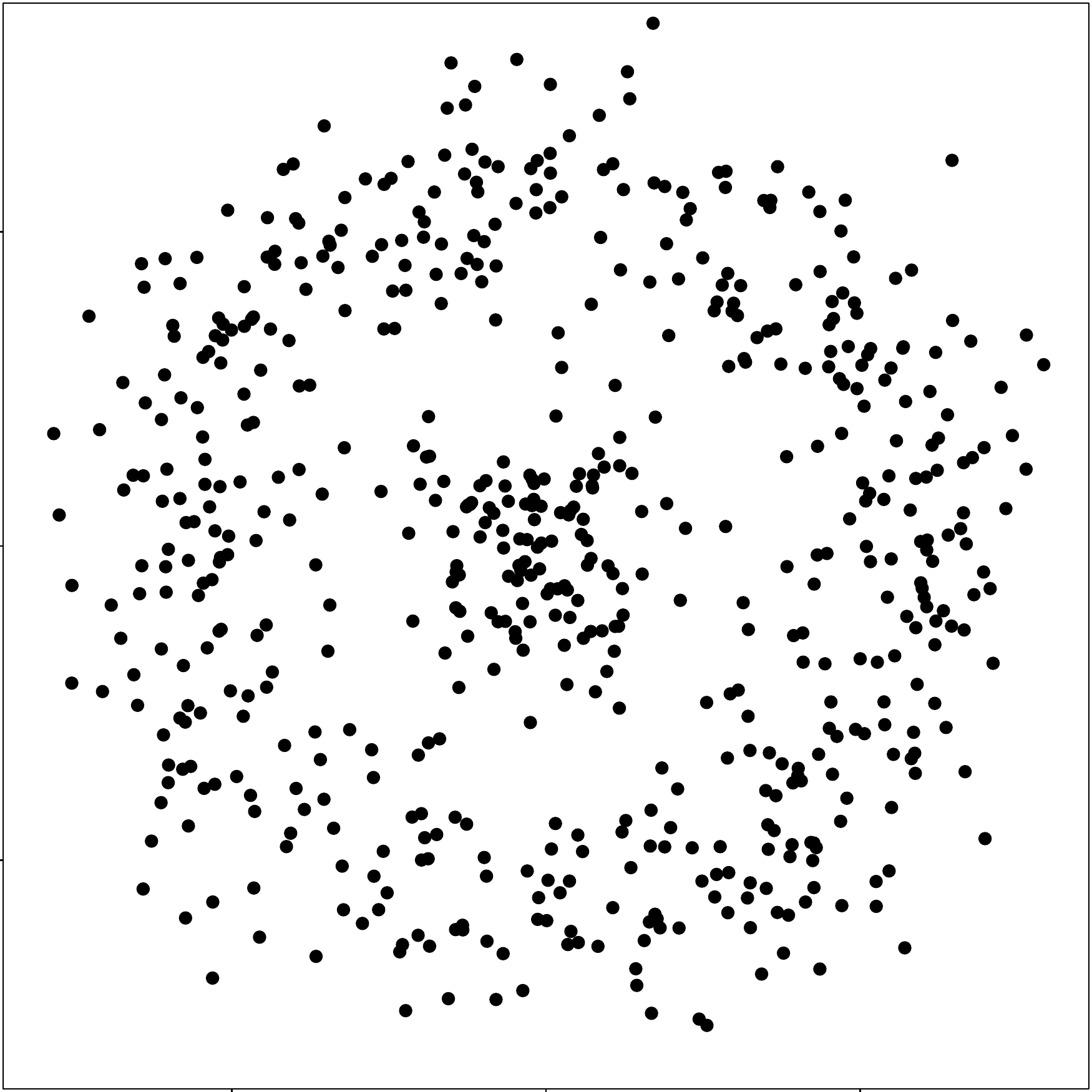}
	\includegraphics[width=1.6in]{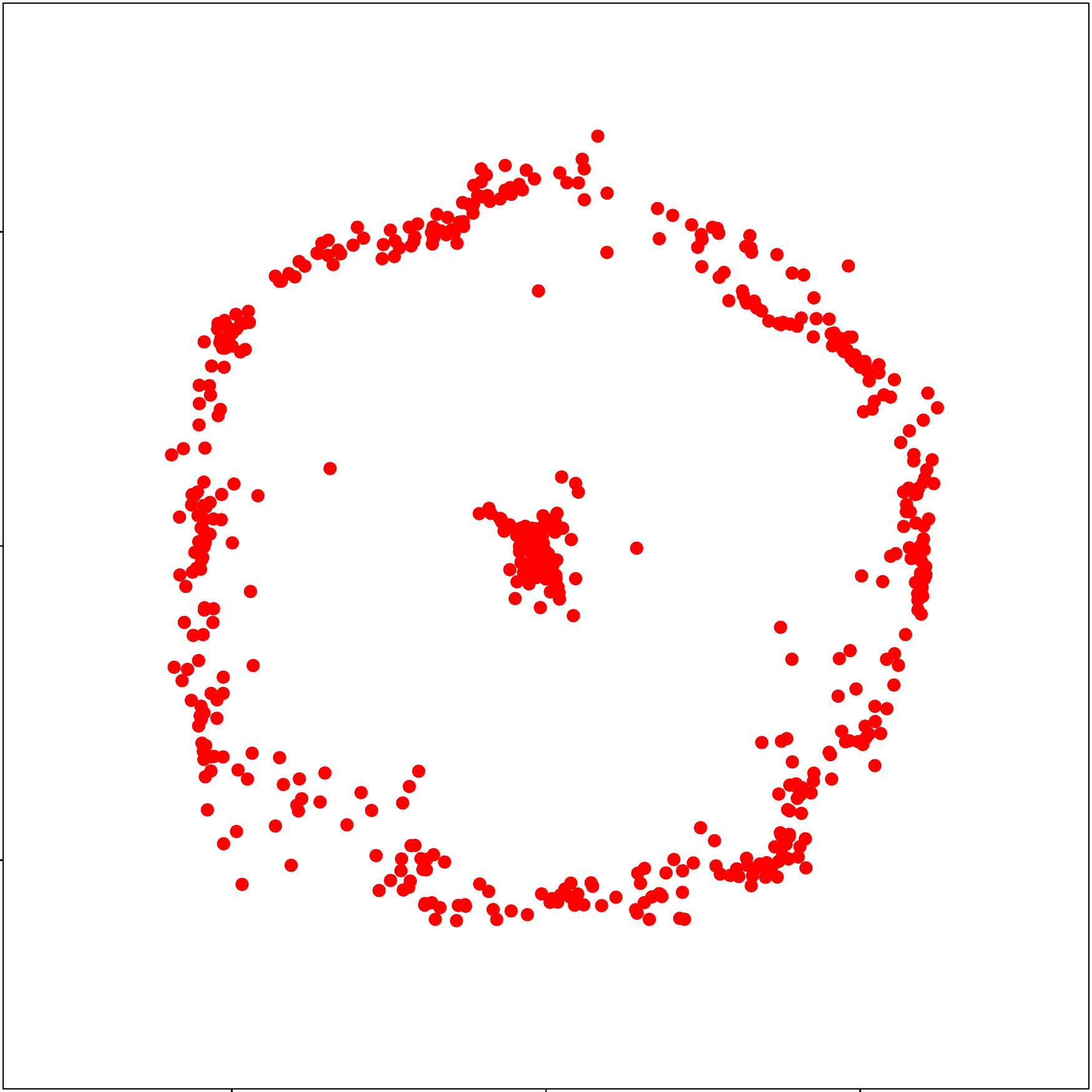}
	\caption{An example of denoising. 
		Left: the original data; there are two structures--an inner small ball region
		and an outer ring area. 
		Right: the denoised data; we apply the mean shift algorithm $3$ times
		to process the data and the structure is now easy to see.}
	\label{fig::ex01}
\end{figure}

(Manifold) Denoising 
is an important task in data analysis \citep{hein2006manifold,wang2010manifold,wang2013sparse} .
The goal of denoising is to pre-process the data so that it is easier to recover the original structure. 
The following is an example of a statistical model that the denoising would help.
We observed IID data points $X_1,\cdots, X_n$ 
from a distribution with density $p(x) = \pi \cdot f(x) + (1-\pi)\cdot u(x)$,
where $0\leq \pi\leq 1$ and $f(x)$ is a density on a structure of interest such as a lower dimensional manifold
and $u(x)$ is a uniform distribution. 
Namely, for each observation, we have some chance to directly observe it from the underlying structure
but there is also some possibility that this observation is a background noise.
In this case, the purpose of denoising is to reduce the effect from background noises.
As an illustrative example, consider Figure \ref{fig::ex01}.
This dataset contains $n=600$ points where with probability $\pi=\frac{5}{6}$ we 
obtain a sample from the actual structure and with probability $1-\pi = \frac{1}{6}$
the observation is from a uniform distribution. 
The actual structure consists of an inner small ball region and an outer ring area. 
The left panel shows the original dataset; we can roughly see the structure.
And the right panel shows the data points after denoising; the structure now becomes clear.

Although there are many other statistical models where the denoising is useful such as the additive model 
(we observe $X=W+\epsilon$, where $W$ is a distribution on the underlying structure and $\epsilon$ is some noise such as a Gaussian),
a common feature of these models 
is that 
the distribution of denoised data points concentrates more around high density areas compared
to the original distribution of the dataset.
Thus, a good denoising approach should reshape the distribution of the data points according to this principle.

In this paper, we propose to use the mean shift algorithm \citep{Fukunaga,Comaniciu,cheng1995mean} to denoise data.
The mean shift algorithm is a popular clustering technique that is widely applied in computer vision and signal processing 
\citep{Fukunaga,Comaniciu,cheng1995mean,carreira2015review}.
The main idea of the mean shift algorithm is to move a given point by taking the gradient ascent
of a density function. 
Because the mean shift algorithm moves a point toward high density areas, 
it is an ideal method for denoising. 

The idea of using the mean shift algorithm as a denoising method has been implemented in 
\cite{Fukunaga,wang2010manifold,carreira2015review}.
However, most of the previous work focused on the analysis of implementations and computational convergence. 
The statistical foundation for how the mean shift works for denoising purposes 
remains unclear
and there is no literature about how the mean shift denoising can improve
statistical analysis.




{\bf Our Contributions.}
We introduce a framework for analyzing how the mean shift algorithm 
is used as a denoising method.
The key element is to view the mean shift algorithm as an operator on
a distribution.
We derive an explicit rate for the concentration of measure around high density regions
and local maxima after applying the mean shift to a distribution. 
We then 
show that by using the mean shift denoising we enhance several
clustering methods, improve the power of two-sample tests, and obtain a new method for anomaly detection.

{\bf Related Work.}
The mean shift algorithm is related to mode clustering \citep{chacon2012clusters,Chacon2014,chen2016comprehensive}.
The statistical analysis for the mean shift algorithm can be found in \cite{arias2016estimation,chen2015statistical,chen2016comprehensive}.
The analysis for the implementations and computations of the mean shift can be found in 
\cite{carreira2003number,carreira2006acceleration,carreira2006fast,carreira2008generalised, wang2007fast}.


{\bf Outline.}
We begin with a short review for the mean shift algorithm in Section~\ref{sec::review} followed by a framework for analyzing the mean shift algorithm as a denoising method in Section~\ref{sec::MSD}.
We analyze how the mean shift algorithm changes an input distribution in Section~\ref{sec::theory}.
We then present three applications of mean shift denoising: an improved clustering, a two-sample test,
and a new anomaly detection in Section~\ref{sec::application}.
Finally we conclude the paper in Section~\ref{sec::discussion}.

\section{Review of the Mean Shift Algorithm}	\label{sec::review}
Let $X_1,\cdots,X_n\in\mathbb{R}^d$ be a random sample from an unknown distribution $P$ with density $p$ supported on a compact set $\K$. 
The mean shift algorithm is a gradient ascent-based algorithm that shifts a given point $x$
by the following updates:
\begin{equation}
x_{\sf new} \longleftarrow \frac{\sum_{i=1}^nX_i K \left(\frac{x-X_i}{h}\right)}{\sum_{i=j}^nK(\frac{x-X_j}{h})},
\label{eq::ms0}
\end{equation}
where the kernel function $K$ is a smooth function
such as the Gaussian $K(u)=(2\pi)^{-d/2}e^{-\|u\|^2/2}$.

Let $\hat{p}_{n,h}(x)$ be the kernel density estimator (KDE):
\begin{equation*}
\hat p_{n,h}(x)=\frac{1}{nh^d}\sum_{i=1}^n K\left(\frac{x-X_i}{h}\right).
\end{equation*}
It is well known \citep{arias2016estimation,cheng1995mean} that the update in equation \eqref{eq::ms0} is equivalent to
\begin{equation}
x_{\sf new} \longleftarrow x + c\cdot h^2\cdot \frac{\nabla \hat{p}_{n,h}(x)}{\hat{p}_{n,h}(x)},
\label{eq::ms2}
\end{equation} 
where $c$ is a known constant.
We call equation \eqref{eq::ms2} the \emph{empirical mean shift}.
An appealing feature for the empirical mean shift is that 
when $K$ is convex and monotonically decreasing, $\hat{p}_{n,h}(x_{\sf new})>\hat{p}_{n,h}(x)$ \citep{Comaniciu}.
Namely, the mean shift algorithm moves a given point into a higher density regions.  
This feature allows us to use the mean shift algorithm as a denoising approach to move data points in the low density areas
into high density areas. 
In particular, when we apply the mean shift algorithm to the data points $X_1,\cdots, X_n$ (choose $x$ to be each of them),
the updating equation \eqref{eq::ms0} shifts data points to higher density regions, which denoises the original data points.
We call this procedure (applying the empirical mean shift algorithm to the data points) the \emph{empirical mean shift denoising (MSD)}.

Because the KDE $\hat{p}_{n,h}$ and its gradient $\nabla \hat{p}_{n,h}$ converges to the population density $p$ and gradient $\nabla p$
under a suitable choice of $h$ \citep{wasserman2006all,scott2009multivariate},
the population version of the updating equation \eqref{eq::ms2} is
\begin{equation}
x^*_{\sf new} \longleftarrow x + c\cdot h^2\cdot \frac{\nabla p(x)}{p(x)}.
\label{eq::ms3}
\end{equation}
We called equation \eqref{eq::ms3} the \emph{population mean shift}
because it is to replace the KDE in the empirical mean shift by the corresponding population quantity.
Later we will see that this population updating equation reveals insights about
how the mean shift algorithm changes a distribution.

\subsection{A Framework for The Mean Shift Denoising}	\label{sec::MSD}

Here we introduce a framework to investigate how the mean shift changes a distribution.
The updating equations \eqref{eq::ms2} and \eqref{eq::ms3}
depend on two quantities: the smoothing bandwidth $h$ and a given density function $p$ (or $\hat{p}_n$).
Thus, we define the operator $\mathbb{M}_{f,\tau}:\K\mapsto \K$ such that for any $x$,
\begin{equation}
\mathbb{M}_{f,\tau}(x) = x + c\cdot \tau^2\cdot \frac{\nabla f(x)}{f(x)},
\label{eq::ms_general}
\end{equation}
where $\tau>0$ a parameter and $f$ is a smooth function. 
Equation \eqref{eq::ms_general} is called the \emph{generalized mean shift.}
It is easy to see that equation \eqref{eq::ms2} corresponds to $\tau=h$ and $f = \hat{p}_{n,h}$
and equation \eqref{eq::ms3} corresponds to $\tau=h$ and $f=p$.
In what follows we discuss about how the three mean shift approaches change a distribution.

\textbf{Empirical Mean Shift.}
In practice, we apply the empirical MSD to denoise a dataset.
This is the case where
we apply the empirical mean shift in \eqref{eq::ms2} to the data points 
$X_1,\cdots, X_n$. This creates
shifted points $\mathbb{M}_{\hat{p}_n,h}(X_1),\cdots, \mathbb{M}_{\hat{p}_n,h}(X_n)$. 
Let $\hat{P}_n$ be the empirical cumulative distribution (eCDF) function of $X_1,\cdots, X_n$
and $\hat{Q}_n$ be the eCDF function of $\mathbb{M}_{\hat{p}_n,h}(X_1),\cdots, \mathbb{M}_{\hat{p}_n,h}(X_n)$. 
Then we can view the empirical mean shift as the operator $\mathbb{M}_{\hat{p}_n,h}$ acting on the eCDF $\hat{P}_n$ that generates a new distribution $\hat Q_n$.
Thus, we write
\begin{equation}
\hat{Q}_n = \mathbb{M}_{\hat{p}_n,h} \otimes \hat{P}_n.
\label{eq::shift_op_hatP}
\end{equation}
This distribution function $\hat{Q}_n$ is a key quantity in our analysis because
it represents the distribution of data points after denoising by the empirical mean shift.

\textbf{Population Mean Shift.}
To understand the difference between $\hat{Q}_n$ and $\hat{P}_n$,
we analyze the counter partners of them--the shifted population distribution $Q$ and the original population distribution $P$.
In more details,
let $X$ be a random variable with distribution $P$ and density $p$, the same as the random sample. 
Let $Q$ be the distribution (with density $q$) of the shift point $\mathbb M_{p,h}(X)$ by equation \eqref{eq::ms3}. 
Then $Q$ and $P$ are linked by 
\begin{equation}
Q = \mathbb{M}_{p, h} \otimes P.
\label{eq::shift_op_P}
\end{equation}
Thus, $Q$ is the population version of $\hat{Q}_n$ and the analysis of the difference between
$Q$ and $P$ reveals information about $\hat{Q}_n$.

\textbf{Generalized Mean Shift.}
The former two cases are applying the mean shift to two distributions which can be casted in
a more general framework using the general mean shift described in
equation \eqref{eq::ms_general}.
Let $X$ be a random variable from a distribution $G$. 
Then the shifted point $\mathbb{M}_{f,\tau}(X)$ has the distribution
\begin{equation}
S_{f,\tau,G} = \mathbb{M}_{f,\tau}\otimes G.
\label{eq::g_shift_op}
\end{equation}
Equation \eqref{eq::g_shift_op} is a general form for analyzing how the mean shift acts on
a distribution. 
It is easy to see that $Q = S_{p,h,P}$ and $\hat{Q}_n = S_{\hat{p}_n,h,\hat{P}_n}$. 
The generalized mean shift provides a flexible framework for analyzing how the mean shift algorithm changes
a distribution. 



\section{Theoretical Results}	\label{sec::theory}


We first define some notation used in describing our theoretical results.
Let $f$ be a function defined on a compact support $\mathbb{K}$.
Define $\|f\|_{0,\infty} = \sup_{x}|f(x)|$, 
$\|f\|_{1,\infty} = \sup_{x}\|\nabla f(x)\|_{\max}$,
and $\|f\|_{2,\infty} = \sup_{x}\|\nabla\nabla f(x)\|_{\max}$ to be the $L_\infty$ norm for
different orders of derivatives.
For a smooth function $f$, 
we say $f$ is a Morse function if its critical points are non-degenerate \citep{morse1925relations,chen2015statistical}.
$f$ being a Morse function is equivalent to saying that the eigenvalues of the Hessian matrix $\nabla\nabla f$ 
at each critical point are non-zero.
We define the (upper) level set of $f$ at level $\lambda$ as
$$
L_{f,\lambda} = \{x: f(x)\geq \lambda\},
$$
which is the regions where the function $f$ is greater than or equal to the level $\lambda$.
Because the level set of the density function $L_{p,\lambda}$ is frequently used in this paper,
for abbreviation we define $L_\lambda = L_{p,\lambda}$.
For any set A, we define $d(x, A)=\inf_{y\in A}\|x-y\|$ be the projection distance from $x$ to $A$.

We consider the following assumptions for a function $f$.
\begin{itemize}
	\item[(A1)] $
	\|f\|_{\ell,\infty}<\infty, \text{ for }\ell=0,1,2
	$
	\item[(A2)] $f$ is a Morse function.
	\item[(A3)] Let $\partial L_{f,\lambda} $ be the boundary of $L_{f,\lambda}$. We assume
	$$
	\inf_{x\in \partial L_{f,\lambda}} \|\nabla f(x)\| \geq g_0>0 .
	$$
	
\end{itemize}
Assumption (A1) is to control the smoothness of the density function.
This is a common assumption for ensuring the stability of both density and gradient estimation 
\citep{chen2015asymptotic,Genovese2010,Genovese2012a}.
Assumption (A2) is a common assumption in level set estimation literature 
\citep{Cadre2006,chen2015density,Tsybakov1997}.
Lower bound on the gradient ensures the stability of level sets.
Morse function (assumption (A3)) is to make sure the population gradient ascent is well-behaved
\citep{azizyan2015risk,chacon2015population,chen2016comprehensive}.

\subsection{Inference for The Population Mean Shift}

To investigate the behavior of the MSD, we start with the analysis of the population mean shift.
Namely, we will first study how the population distribution $P$ changes after denoising. 
The following two theorems show the difference between $P$ and $Q$ 
(and the corresponding densities $p$ and $q$), providing us an intuition about how the mean shift algorithm serves as a denoising process. 
\begin{theorem}
	Assume the density function $p$ satisfies (A1--2). 
	Then for a level set $L_{\lambda}$ satisfing (A3),
	if the bandwidth $h^2 \leq \min\left\{\frac{3\sqrt{2}\lambda}{c\cdot\|f\|_{2,\infty}}, \frac{\sqrt{2}\lambda^2}{c\cdot g_0^2}\right\}$, 
	the probability mass within the upper level set $L_\lambda$ after denoising will at least increase by
	$$
	Q(L_\lambda) - P(L_\lambda) \geq 
	c\cdot h^2 \frac{g_0}{6\sqrt{2}\lambda}\cdot {\sf Vol}_{d-1}(\partial L_\lambda),
	$$
	where ${\sf Vol}_{d-1}(A)$ is the $(d-1)$-dimensional hypervolume of set $A$. 
	Namely, $Q(L_\lambda) - P(L_\lambda) = O(h^2)$ as $h\rightarrow0$.
	\label{thm::QP}
\end{theorem}
Intuitively, the probability content would concentrate more inside the level set after shifted by the population mean shift.
The feature of Theorem \ref{thm::QP} is that it quantifies the increasing rate of probability when $h$ is small.

\begin{theorem}
	Assume density $p$ satisfies (A1--2). 
	\begin{itemize}
		\item If $m$ is a local mode of $p$, then when $h^2<\frac{p(m)}{c\cdot \|p\|_{2,\max}}$, there exists positive constants $c_1,c_2$ such that
		$$
		0<c_1 h^2\leq \frac{q(m)}{p(m)}-1 \leq c_2 h^2.
		$$
		\item If $m$ is a local minimum of $p$ and $p(m)>0$, then when $h^2<\frac{p(m)}{c\cdot \|p\|_{2,\max}}$, there exists positive constants $c_1,c_2$ such that
		$$
		0<c_1 h^2\leq 1-\frac{q(m)}{p(m)} \leq c_2 h^2.
		$$
	\end{itemize}
	\label{thm::mode}
\end{theorem}
Theorem~\ref{thm::mode} quantifies the increase/decrease in the density at local modes/minima after applying the (population) mean shift algorithm. 
It is expected that $q(m)>p(m)$ at local modes and $q(m)<p(m)$ at local minima due to the nature of gradient ascent, but here we further obtain the rate of increment $O(h^2)$.

\subsection{Inference for The Empirical Mean Shift}

Now we turn to the empirical mean shift \eqref{eq::shift_op_hatP} and 
study the behavior of operator $\mathbb{M}_{\hat {p}_{n,h}, h}$ on the empirical CDF $\hat{P}_n$.
We will show that the distribution $\hat{Q}_n$ of shifted data points $X'$ concentrates around high density regions of the distribution $P$.  To see this, we first introduce an intermediate distribution
$$
\overline{Q}_n = \mathbb{M}_{\hat p,h} \otimes P,
$$
which is the distribution after applying the empirical mean shift algorithm $\mathbb{M}_{\hat {p}_{n,h}, h}$ to the true distribution $P$. The difference between $\overline{Q}_n$ and $Q$ provides information about how the empirical mean shift $\mathbb{M}_{\hat p,h}$ and the population mean shift $\mathbb{M}_{p,h}$ differ from each other. The difference between $\overline{Q}_n$ and $\hat{Q}_n$ can further tell us about how the difference in pre-shifted distributions $P$ and $\hat P$ results in the difference of post-shifted distributions by implementing emprical mean shift $\mathbb{M}_{\hat p,h}$.
\begin{theorem}
	Let $\mathbb{K}$ be the support of $P$, assume $p$ satisfies (A1--2) and $p(x)\geq p_0 >0, ~\forall x\in \K$. 
	Let $\delta_{1,n} = \max\{\|\hat{p}_n-p\|_{\ell,\infty}: \ell=0,1\}$.	
	Then for any given set $A\subset \K$, 
	$$
	\overline{Q}_n(A) - Q(A) = O_P(h^2\cdot \delta_{1,n}), \text{ as } \delta_{1,n}\rightarrow 0.
	$$
	\label{thm::QQ}
\end{theorem}
Theorem~\ref{thm::QQ} measures the difference in distributions after applying the population mean shift and the empirical mean shift.
This is somewhat expected since the difference in two MSD approaches is in the quantity $\frac{\nabla p(x)}{p(x)}$ and $\frac{\nabla \hat p_{n,h}(x)}{\hat p_{n,h}(x)}$.
Therefore, as long as both gradient and density are similar, the shifted position by both approaches should be close to each other. 
Note that under smoothness assumptions on the kernel function, 
the rate of $\delta_{1,n} = O(h^2) + O_P\left(\sqrt{\frac{\log n}{nh^{d+2}}}\right)$;
see, e.g., \cite{genovese2014nonparametric} and \cite{chen2015statistical}.

The following theorem investigates how the difference between pre-shifted distribution $P$ and $\hat P$ contributes to 
the difference in the post-shifted distributions shifted by empirical MSD.
\begin{theorem}
	For any given set $A$, 
	$$
	\hat{Q}_n(A) - \overline{Q}_n(A)  = O_P\left(\sqrt{\frac{1}{n}}\right).
	$$
	\label{thm::estimate}
\end{theorem}
Theorem~\ref{thm::estimate} is reasonable since
the difference between $\hat{Q}_n$ and $\overline{Q}_n$ results from the difference between the pre-shifted distributions: $\hat{Q}_n$
is shifted from the empirical cumulative distribution function(CDF) $\hat P$ while $\overline{Q}_n$ is from the true CDF $P$.
And it is well known that for a given set $A$, the empirical CDF is a rate $O_P\left(\sqrt{\frac{1}{n}}\right)$ estimator
to the population CDF.

Thus, by putting all theorems together, we have the following result.
\begin{corollary}
	Assume $p$ satisfies (A1--2) with $p(x)\geq p_0, \forall x\in\K$. Then we have
	\begin{align*}
	\hat{Q}_n(L_\lambda) - P(L_\lambda) &=O(h^2)+O_P\left(\sqrt{\frac{1}{n}}\right) + O(h^2\cdot \delta_{1,n}),\\
	\hat{Q}_n(\hat{L}_\lambda) - P(\hat{L}_\lambda) &=O(h^2)+O_P\left(\sqrt{\frac{1}{n}}\right) + O(h^2\cdot \delta_{1,n}),
	\end{align*}
	where $\hat L_{\lambda} = L_{\hat{p}_n,\lambda}$.
	\label{cor::LV}
\end{corollary}
Note that the first quantity
$O(h^2)$ in Corollary \ref{cor::LV} is always positive and it represents the increase in probability measure at $L_\lambda$ 
(see Theorem~\ref{thm::QP}). 
A good news is that it is
the dominating term among the three quantities in the Corollary \ref{cor::LV}. 
Therefore, when we apply mean shift algorithm to a random sample, the shifted distribution does concentrate more on the high density regions of the population.

\subsection{Inference for The Generalized Mean Shift}

As shown in equation \eqref{eq::ms_general}, the mean shift procedure can be generalized to any smooth function $f$. Thus, we derive the following theorem that shows the difference between the post-shifted distribution $S=S_{f,\tau,P}$ and the pre-shifted $P$. 
Let $s$ be the density function of $S$ and $p$ be the density function of $P$.
\begin{theorem}
	For any function $f$ satisfing (A1--2). Let $\lambda_{\min}$ be the minimal absolute eigenvalue of all critical points of $f$. Then
	\begin{itemize}
		\item \textbf{Density at local modes:} If $m$ is a local mode of $f$, when $\tau^2<\frac{2f(m)}{c\cdot \lambda_{\min}}$, there exists positive constants $c_1$ and $c_2$ that only depends on $f$ such that
		$$
		0<c_1 \tau^2\leq \frac{s(m)}{p(m)}-1 \leq c_2 \tau^2.
		$$
		That is $s(m) - p(m) = O(\tau^2)$.
		\item \textbf{Density at Local minima:} If $m$ is a local minimum of $f$ with  $p(m)>0$, when  $\tau^2<\frac{2f(m)}{c\cdot \lambda_{\min}}$, then there exists positive constants $c_1,c_2$ that only depends on $f$, such that
		$$
		0<c_1 \tau^2\leq 1-\frac{s(m)}{p(m)} \leq c_2 \tau^2.
		$$
		That is $p(m) - s(m) = O(\tau^2)$.
		\item \textbf{Probability mass within level sets:}  Assume $L_{f,\lambda}$ satisfies (A3). Assume that there exists constants $\epsilon_0,\rho_0>0$ such that
		$$
		\inf_{x\notin L_{f,\lambda}, d(x,L_{f,\lambda})\leq \epsilon_0} p(x)  \geq \rho_0.
		$$
		Then
		$$
		S(L_\lambda) - P(L_\lambda) \geq 
		\rho_0\cdot c\cdot \tau^2 \frac{g_0}{3\sqrt{2}\lambda} \cdot {\sf Vol}_{d-1}(\partial L_\lambda).
		$$
		Namely, $S(L_\lambda) - P(L_\lambda) = O(\tau^2)$ as $\tau\rightarrow0$. 
	\end{itemize}
	\label{thm::g_mode}
\end{theorem}
Theorem~\ref{thm::QP} and \ref{thm::mode} are special cases of Theorem~\ref{thm::g_mode}
by identifying $f=p$ and $S=Q$.
An interesting result from Theorem~\ref{thm::g_mode} is that the shifting process in \eqref{eq::ms_general}
concentrates the distribution around high density areas of $f$, regardless of the density of the pre-shifted distribution $P$.
This is because the shifting operation $\mathbb{M}_{f,\tau}$ depends only on $f$.

Moreover, Theorem~\ref{thm::g_mode} can be used to analyze the case where we apply the mean shift algorithm multiple times. For instance, consider the case where we implement the empirical mean shift algorithm to the population distribution $P$. Let $m$ be a local mode of $\hat{p}_{n,h}$ and $\overline{q}_n^{(N)}$ be the density function after being shifted $N$ times. Then by Theorem \ref{thm::g_mode},
$$
\overline{q}_n^{(N)}(m) \geq p(m)\left(1+c_1 h^2\right)^N,
$$
where $c_1>0$ is some constant depending only on $\hat{p}_{n,h}$. This shows the rate of concentration of probability measure around local modes of $\hat{p}_n$ after applying the mean shift algorithm multiple times.

Finally, we derive a perturbation theorem for the shifted distribution $S_{f,\tau,P}$ to investigate how it varies when we slightly perturb each component $f$, $\tau$, and $P$. 
\begin{theorem}
	Assume $f$ satisfy (A1--2), and $\|p\|_{0,\infty}<\infty$. 
	Let $\K$ be the support of $P$. 
	We assume $f(x)\geq f_0>0$, $\forall x\in \K$.
	Then for a given set $A \subset \mathbb{K}$,
	\begin{itemize}
		\item \textbf{Situation 1:} for any sequence $\{f_n\}$ such that $\Delta_{1,n} = \max\{\|f_n-f\|_{\ell,\infty}: \ell=0,1\}\rightarrow 0$, then
		\begin{align*}
		S_{f_n,\tau,P}(A)-S_{f,\tau,P}(A) = O(\tau^2\cdot \Delta_{1,n}).
		\end{align*}
		\item \textbf{Situation 2:} for any sequence $\{\tau_n\}$ such that $|\tau_n-\tau|\rightarrow 0$, then
		\begin{align*}
		S_{f,\tau_n,P}(A)-S_{f,\tau,P}(A) = O(|\tau_n-\tau|).
		\end{align*}
		\item \textbf{Situation 3:} for any sequence $\{P_n\}$ such that $|P_n(B)-P(B)|\rightarrow 0$ where $B$ is any given set, then
		\begin{align*}
		S_{f,\tau,P_n}(A)-S_{f,\tau,P}(A) = O(|P_n(B)-P(B)|).
		\end{align*}
	\end{itemize}
	\label{thm::perturb}
\end{theorem}
Theorem~\ref{thm::perturb} is a perturbation theorem for the shifted distribution $S_{f,\tau,P}$.
It shows that under a small perturbation of each quantity $f$, $\tau$, or $P$, 
the shifted distribution changes linearly with respect to the perturbation.
One can also view Theorem~\ref{thm::perturb} as a generalized Lipchitz property under different metric spaces.
Note that Theorem~\ref{thm::QQ} and \ref{thm::estimate} are both special cases of Theorem~\ref{thm::perturb}.

\section{Applications}	\label{sec::application}

To show how the MSD would help statistical analysis,
we consider three statistical tasks:
(i) clustering, (ii) two-sample test and (iii) anomaly detection.
Note that to apply the empirical mean shift, we need to choose the smoothing bandwidth $h$.
Here we use the smoothed cross validation (SCV) approach \citep{chacon2011asymptotics, chen2016comprehensive}, 
which is based on the approximation of mean integrated square error (AMISE).

\begin{figure*}[h]
	\vspace{.3in}
	\centering
	\begin{subfigure}[t]{\dimexpr0.16\textwidth+10pt\relax}
		\makebox[10pt]{\raisebox{40pt}{\rotatebox[origin=c]{90}{\text{Before MSD}}}}%
		\includegraphics[width=1\dimexpr\linewidth-10pt\relax]{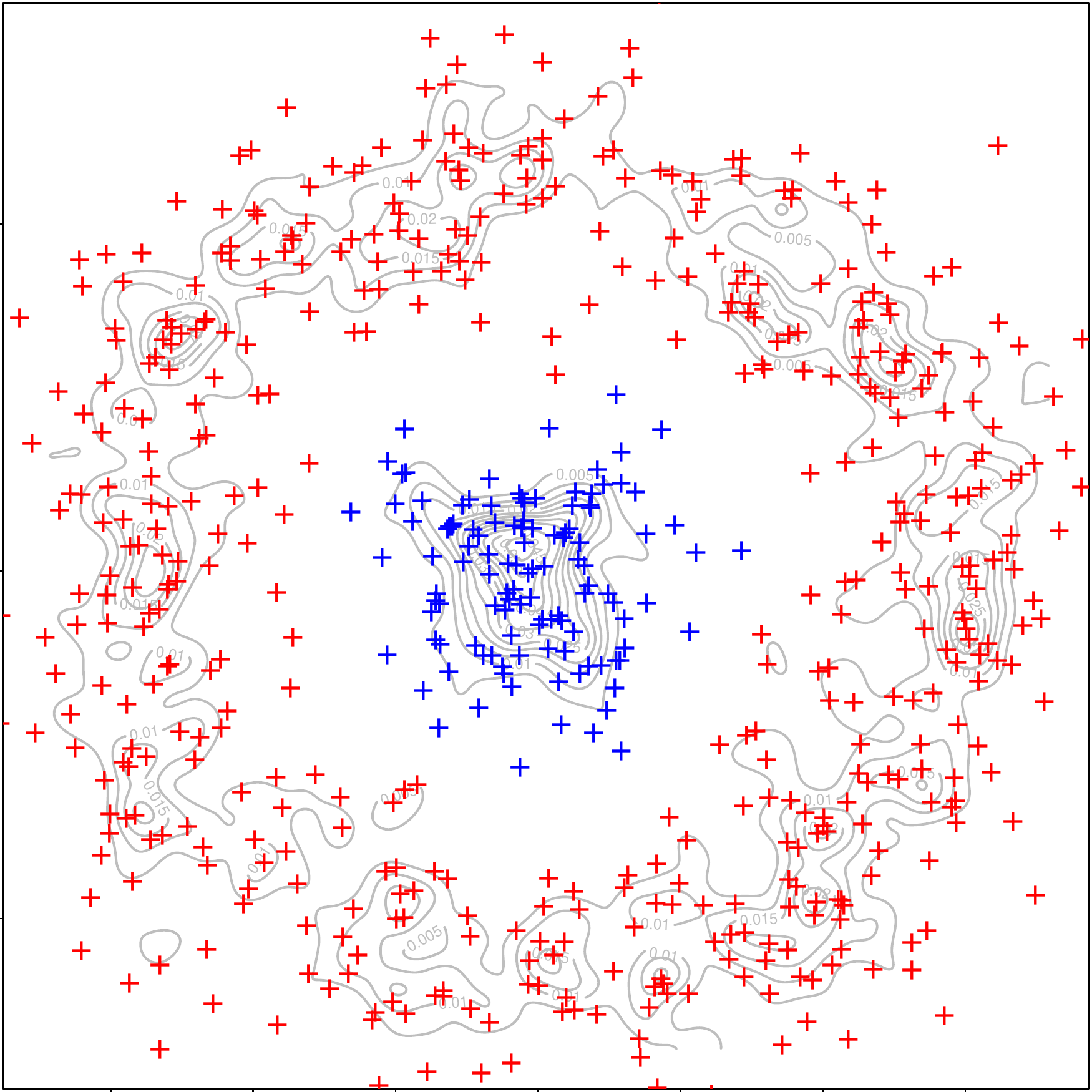}
		\makebox[10pt]{\raisebox{40pt}{\rotatebox[origin=c]{90}{\text{After MSD}}}}%
		\includegraphics[width=1\dimexpr\linewidth-10pt\relax]{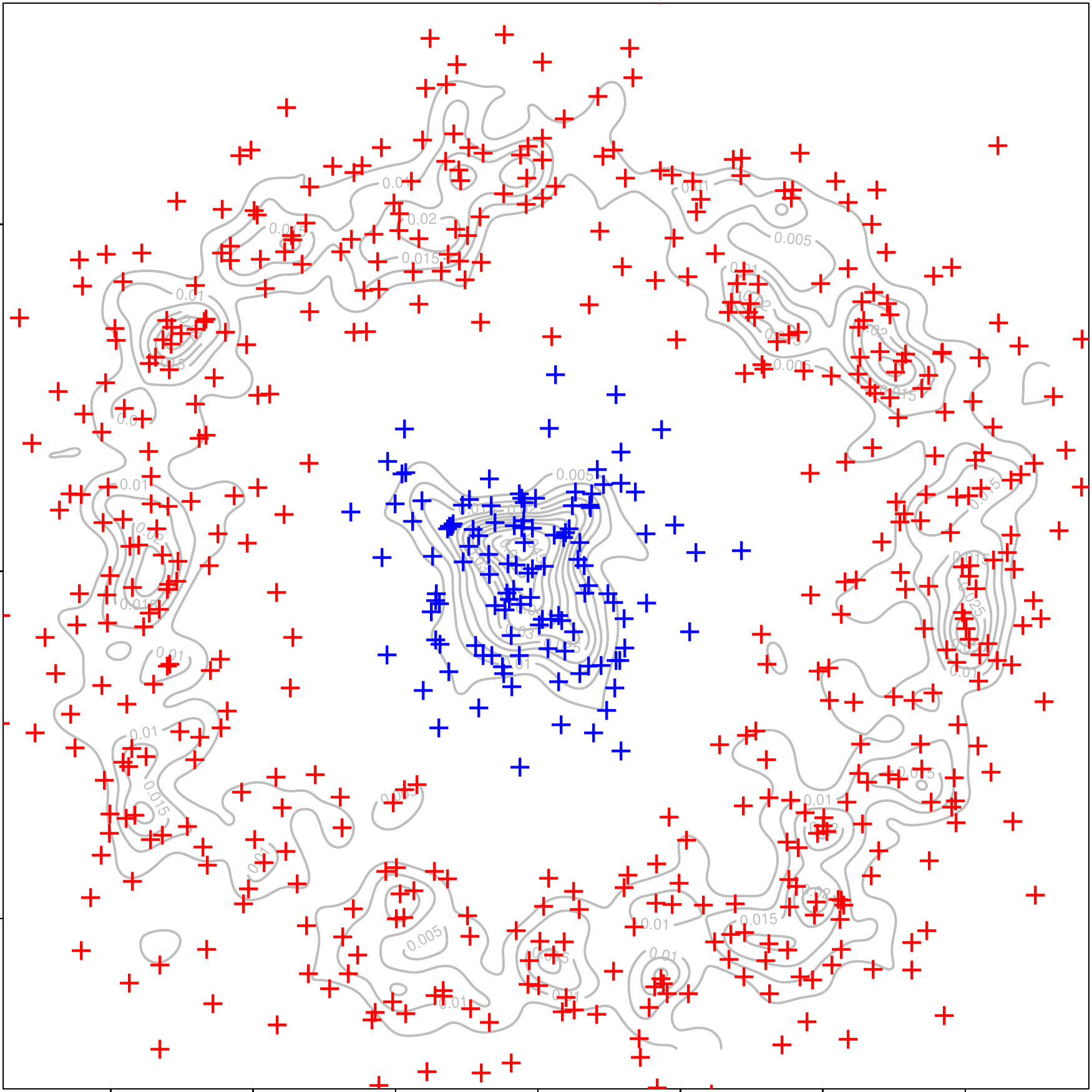}
		\caption{Case 1}
		\label{fig::data_unifnoise1}
	\end{subfigure}%
	\begin{subfigure}[t]{0.16\textwidth}
		\includegraphics[width=1\textwidth]  {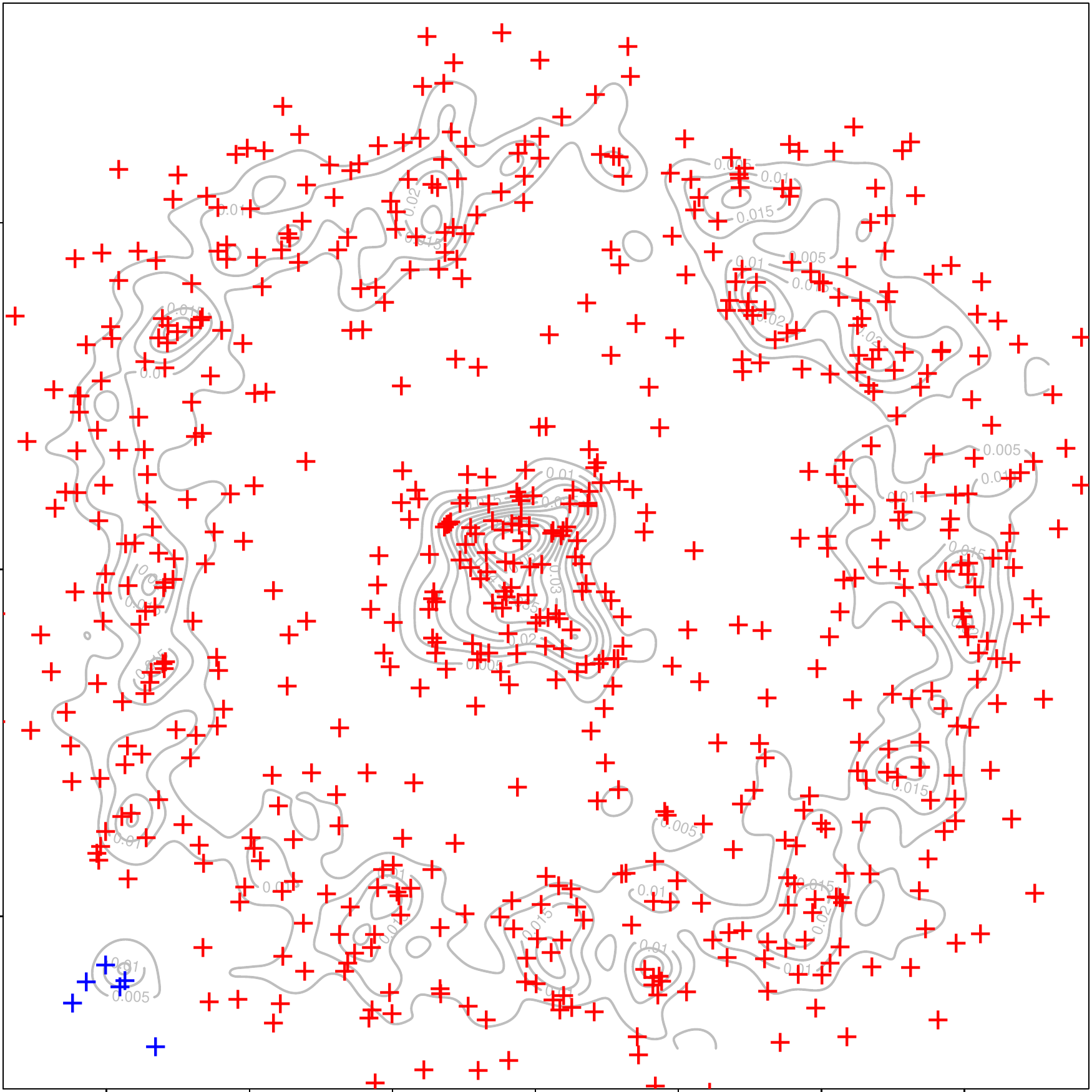}
		\includegraphics[width=1\textwidth]{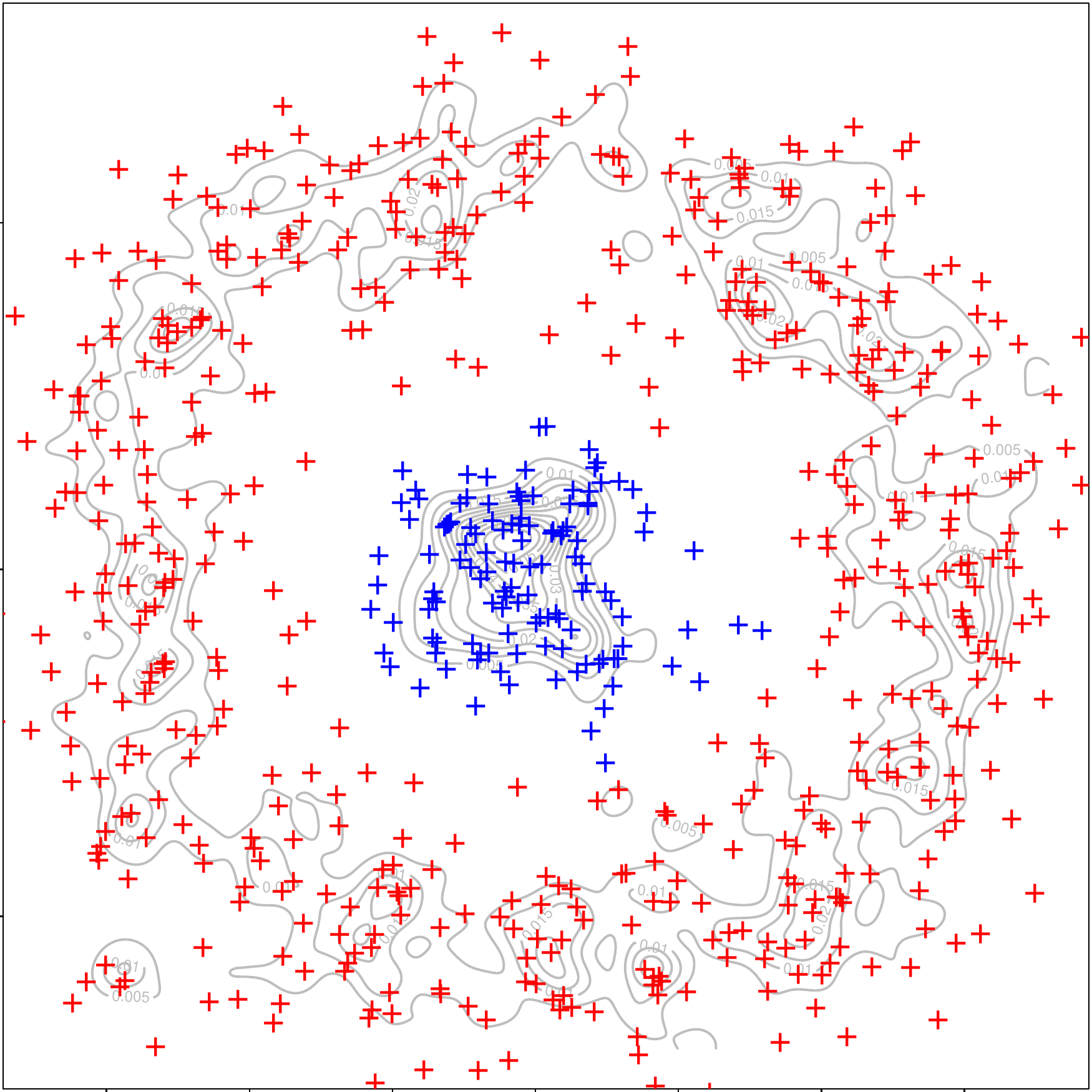}
		\caption{Case 2}
		\label{fig::data_unifnoise2}
	\end{subfigure}%
	\begin{subfigure}[t]{0.16\textwidth}
		\includegraphics[width=1\textwidth] {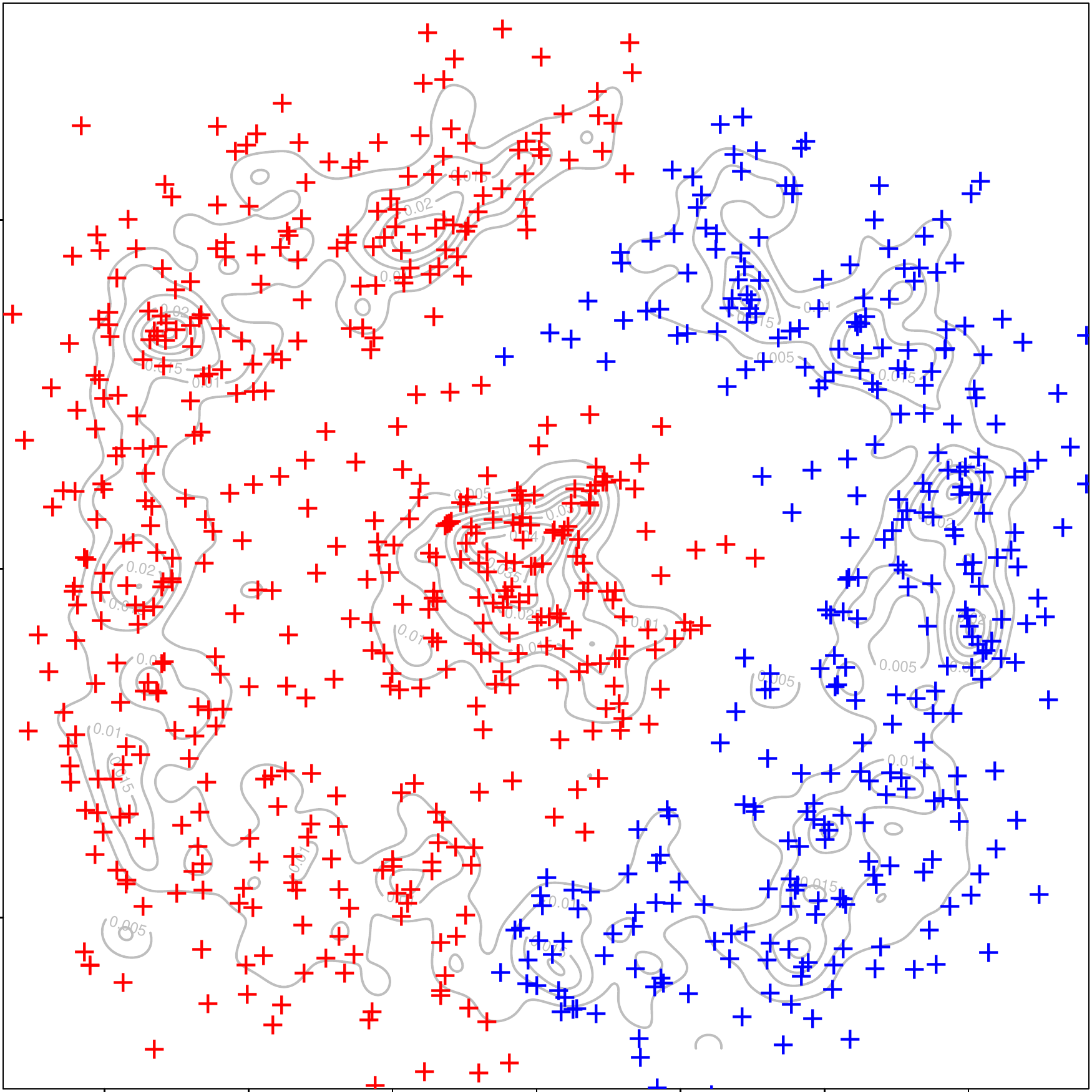}
		\includegraphics[width=1\textwidth]{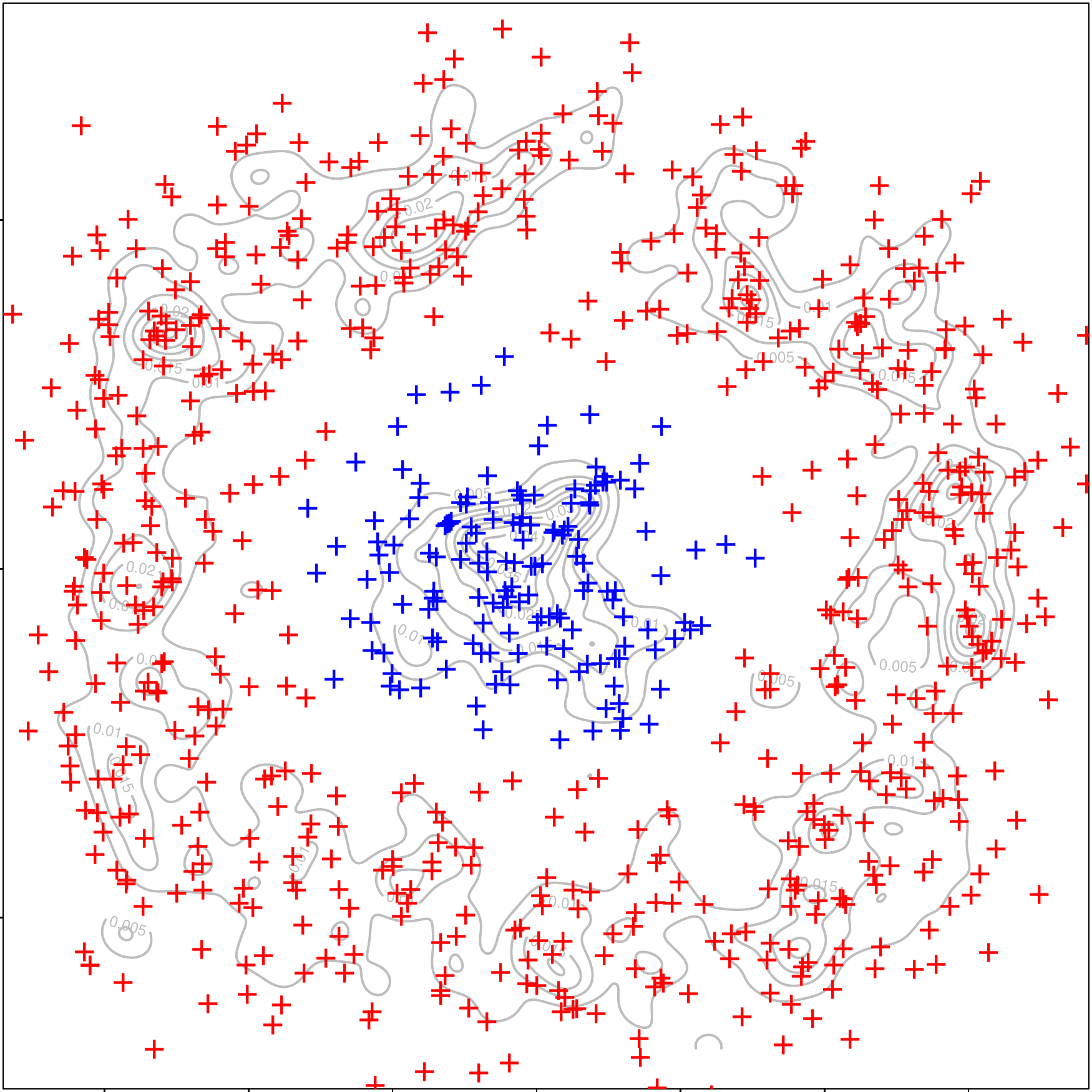}
		\caption{Case 3}
		\label{fig::data_unifnoise3}
	\end{subfigure}%
	\begin{subfigure}[t]{0.16\textwidth}
		\includegraphics[width=1\textwidth]  {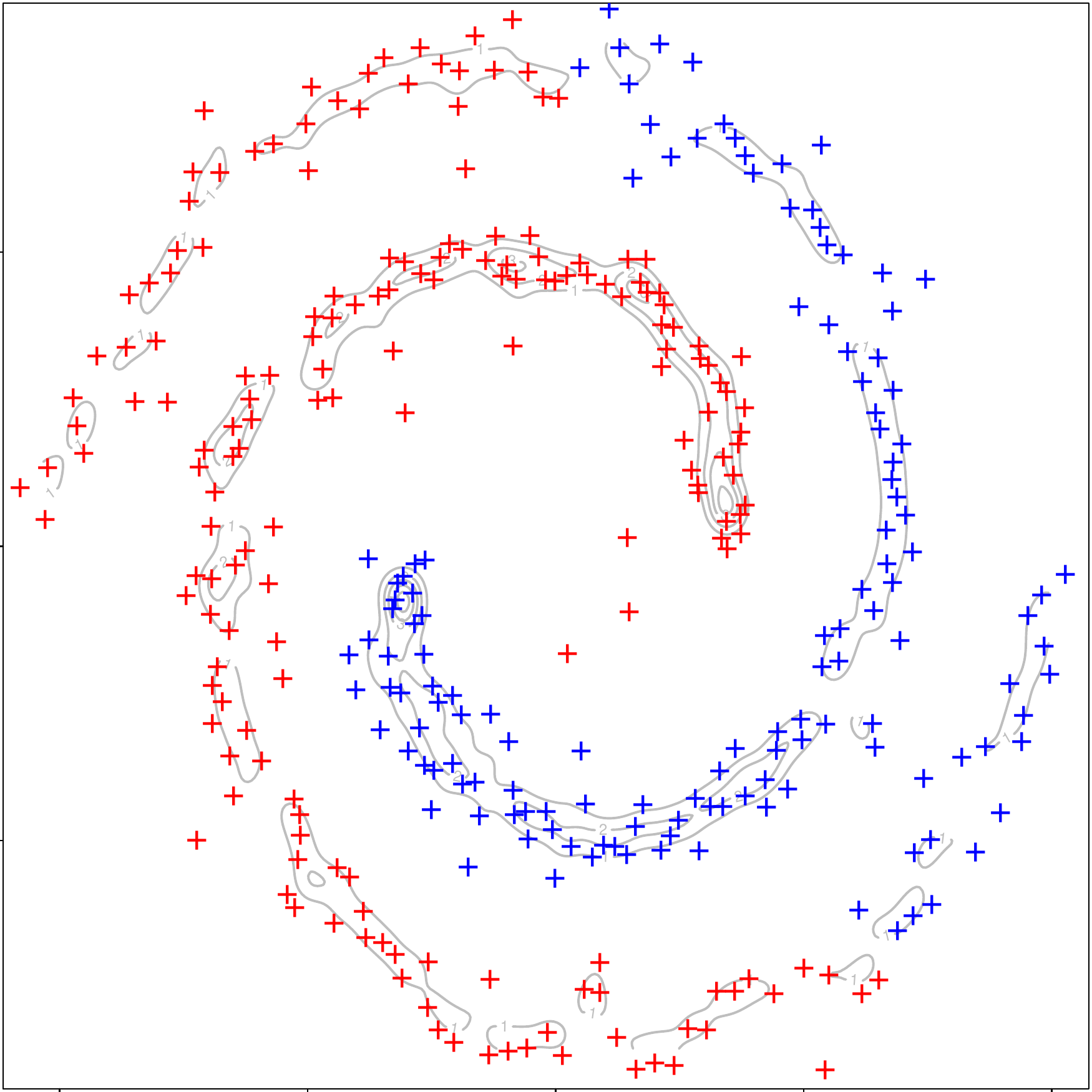}
		\includegraphics[width=1\textwidth]{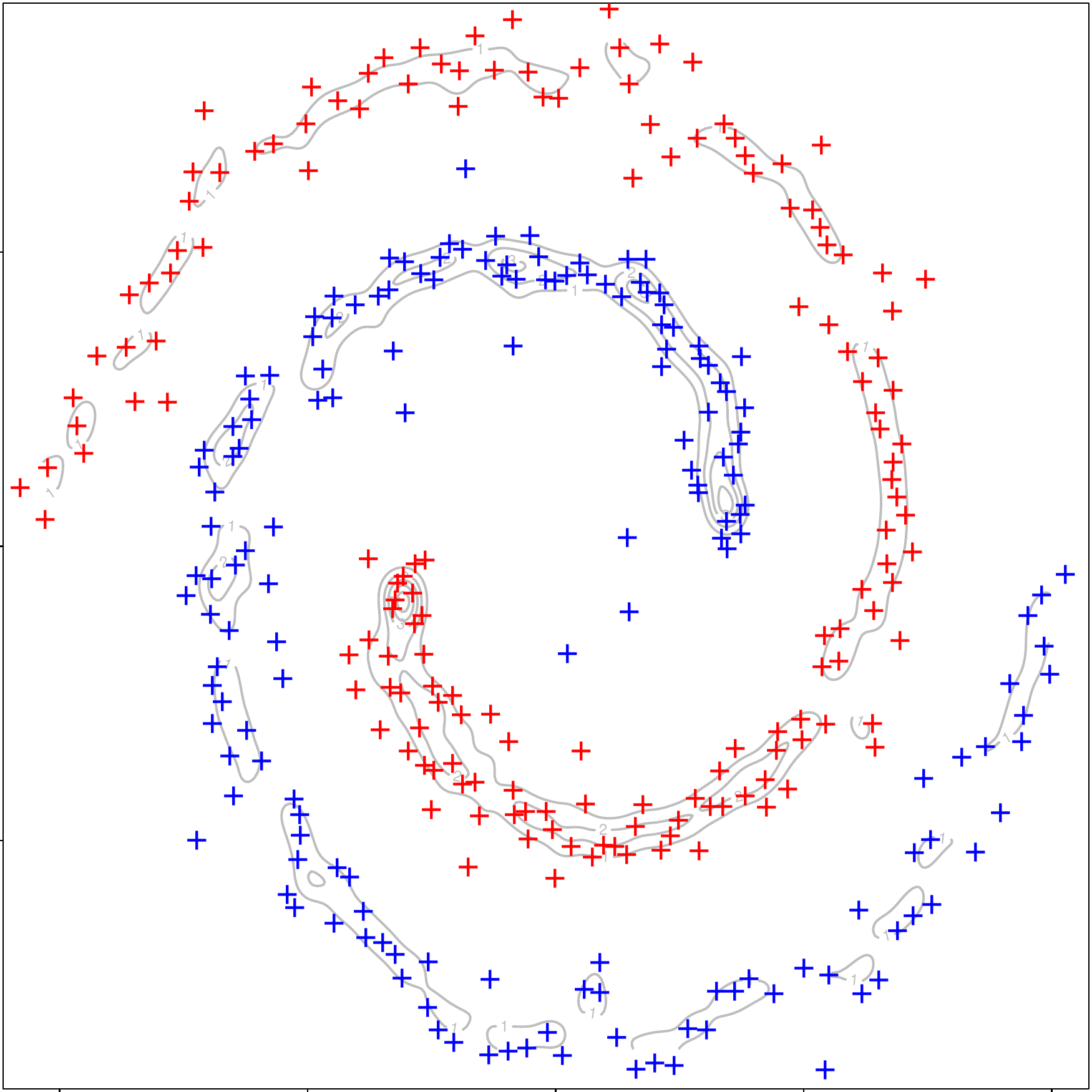}
		\caption{Case 4}
		\label{fig::data_unifnoise4}
	\end{subfigure}%
	\begin{subfigure}[t]{0.16\textwidth}
		\includegraphics[width=1\textwidth]  {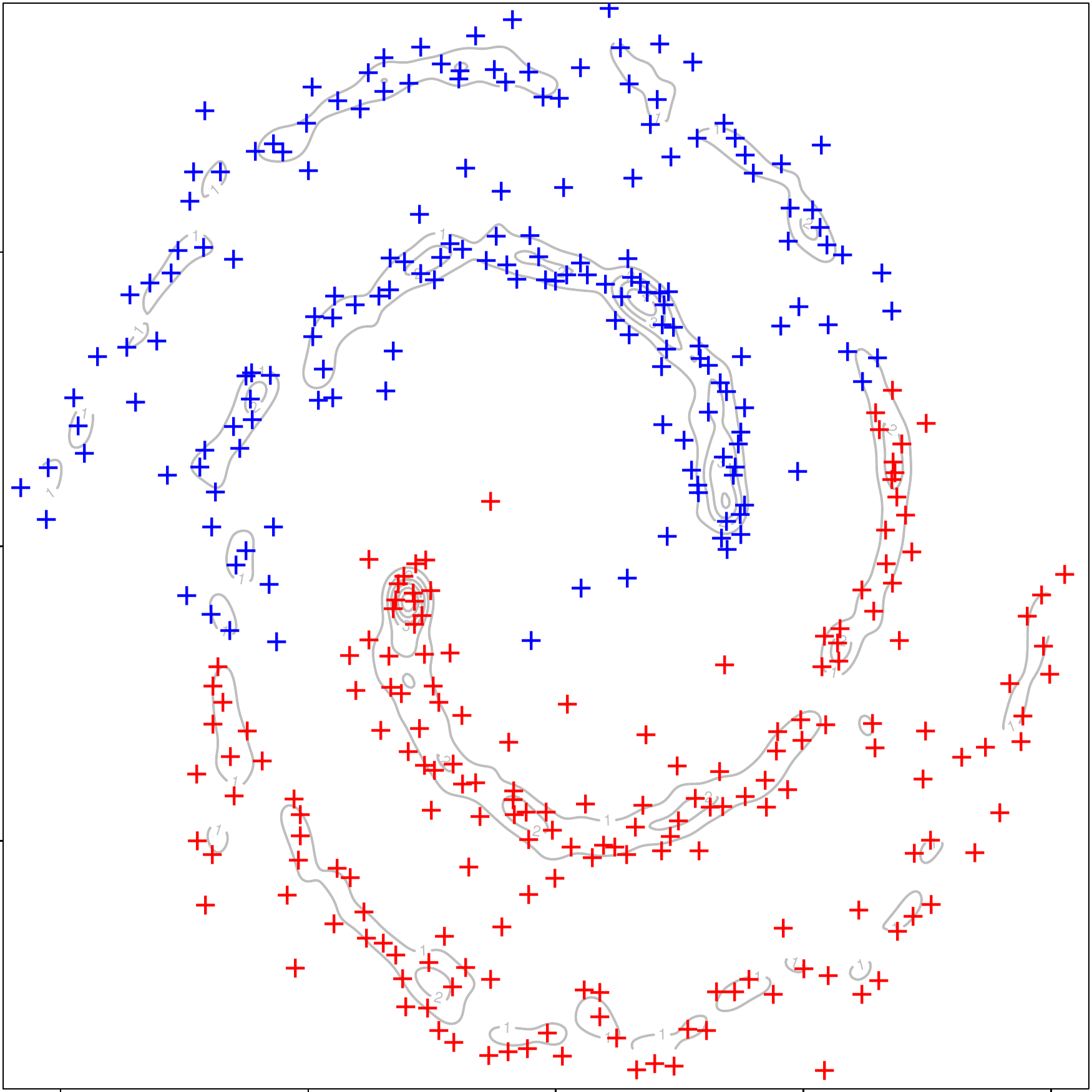}
		\includegraphics[width=1\textwidth]{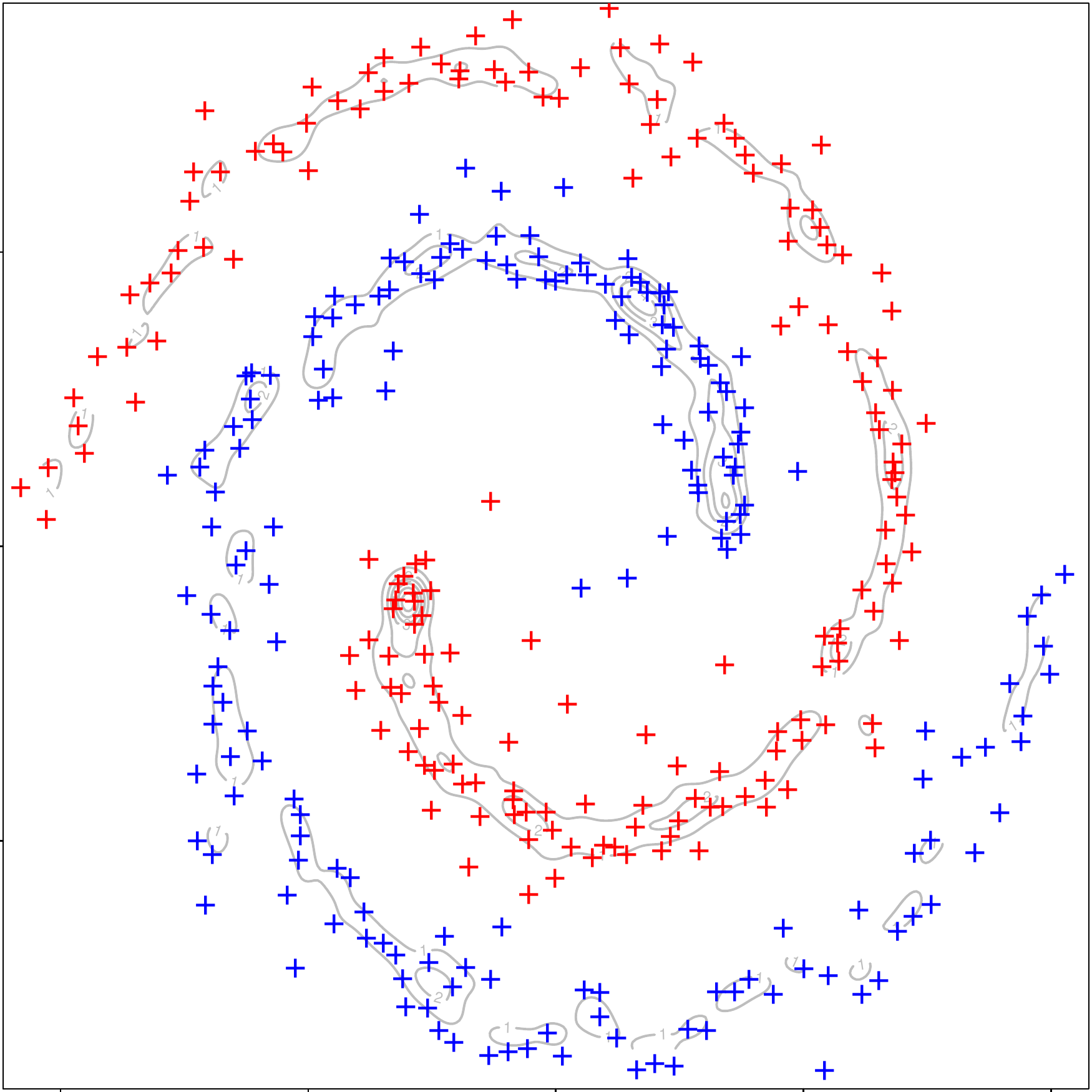}
		\caption{Case 5}
		\label{fig::data_unifnoise5}
	\end{subfigure}%
	\begin{subfigure}[t]{0.16\textwidth}
		\includegraphics[width=1\textwidth] {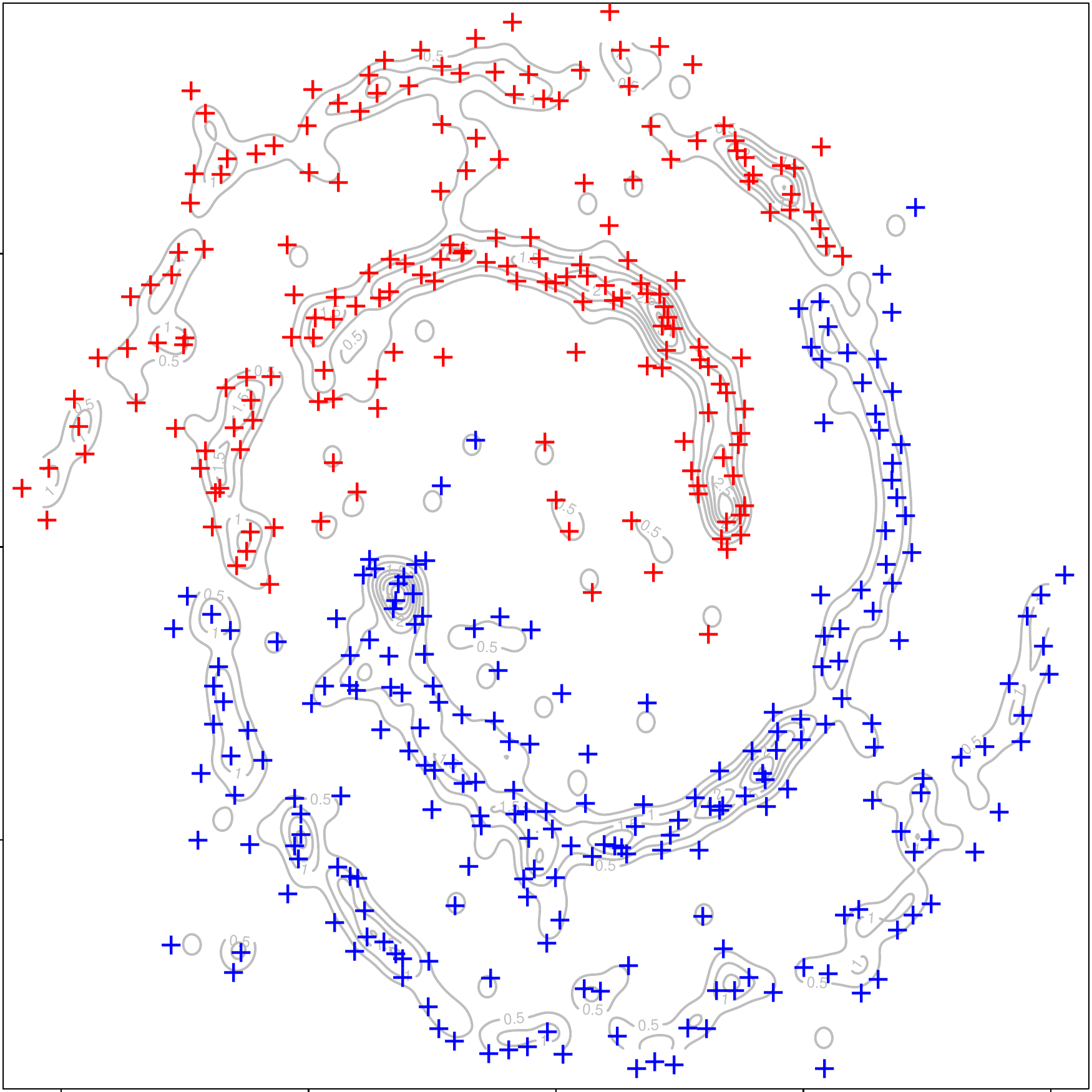}
		\includegraphics[width=1\textwidth]{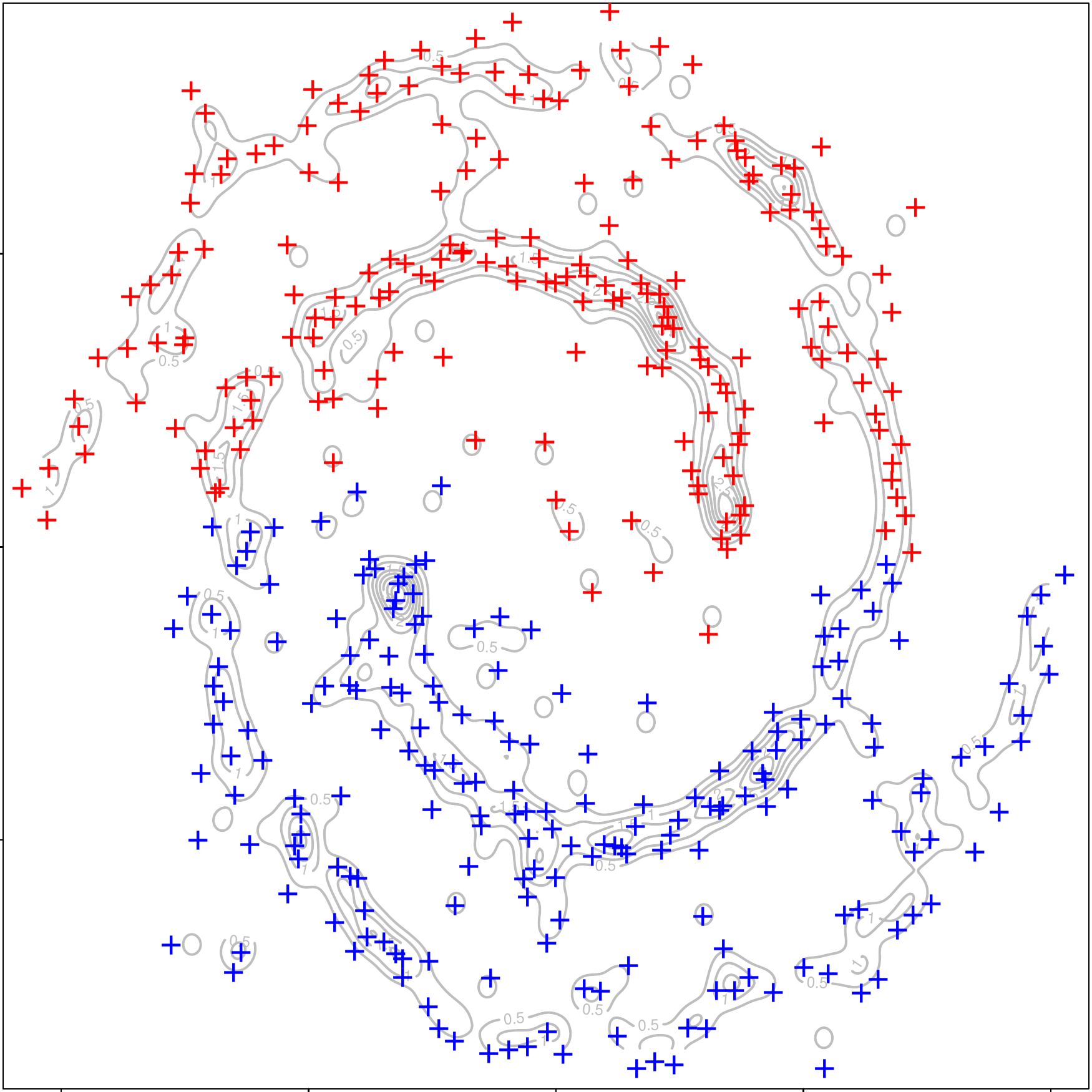}
		\caption{Case 6}
		\label{fig::data_unifnoise6}
	\end{subfigure}
	\caption{Simulated datasets with uniform noises. The first row represents the results of spectral clustering before denoising while the second row shows the results after denoising. Case 1-3 are Bullseye data in which $N_0=500$, the noises are added with an increasing size $N_1=100,150,300$, the clustering results in these cases are improved by MSD. Case 4-6 are for spiral structure and we let $(N_0:N_1)=(300:20), (300:50), (300:100)$ for each case. The denoising in this pattern is easy to fail when the noise ratio is large}
	\label{fig::data_unifnoise}
\end{figure*}

\subsection{Enhancements in Clustering}

Here we consider two simulated and three real datasets. 
In all the cases, we apply the empirical mean shift for only one time
and compare the clustering performance under the pre-denoised versus the post-denoised datasets.
To evaluate the performance of clustering, we use the Adjusted Rand Index (ARI) \citep{hubert1985comparing, steinley2004properties}.

\subsubsection{Simulated Data}

Each simulated data is generated as follows. 
We first generate $N_0$ data points from a distribution on the actual structures and then generate $N_1$ points
from the background noises.
We demonstrate how spectral clustering can be improved by the MSD. 

\textbf{Bullseye Data (Case 1-3).} 
The true structure has a bullseye structure, which consists of a ring with radius $r$ and a central eye, where the eye's fraction is $\pi$. 
We uniformly generate points on the ring and at the center of the central eye and then add a Gaussian noise
with a standard deviation $\sigma$. 
We choose $r=6, \pi=0.2, \sigma=1$ and $N_0=500$ in all cases (Case 1-3).

The background noises are generated from a 2D uniform distribution in the square $[-6.5,6.5]^2$. 
The sample size $N_1$ of Case 1-3 is $N_1=100,150,300$. 

We plot one result of each case for illustration. 
When the data is noisy (Case 2 and Case 3), spectral clustering fails to recover the actual partition
but when we pre-process the data using the MSD, the spectral clustering works.

In Table \ref{tab::be}, we summarize the mean and the standard deviation of the ARI from 200 repetitions for each of Case 1-3. As can be seen, the mean of the ARI after the MSD is much higher than the clustering performance without the MSD. 
This provides evidence that the MSD improves the performance of clustering. 

\begin{table}[h]
	\centering
	\begin{tabular}{llll}
		\hline
		{\bf }  &{\bf Case 1} & {\bf Case 2} &{\bf Case 3} \\
		\hline
		Before &  0.617(0.450)  &  0.452(0.453)  &  0.280(0.389) \\
		After   &  0.954(0.160)  &  0.902(0.262)  &  0.880(0.291) \\
		\hline
	\end{tabular}
	\caption{Adjusted Rand Index for Bullseye Datasets.}
	\label{tab::be}
\end{table}

\textbf{Spiral Data (Case 4-6).} We generate $N_0=300$ data points with a true structure of 2 spirals and add Gaussian noise with $\sigma=0.05$ for each of Case 4-6. The background noises are generated from a 2D uniform distribution in square $[-0.8,0.8]^2$. The sample size of noises in Case 4-6 is $N_1=20, 50, 100$.

We give one example of each case in Figure \ref{fig::data_unifnoise4}-\ref{fig::data_unifnoise6}. It is harder for spectral clustering to discover the clusters under the spiral structure. For Case 4 and Case 5, spectral clustering works after denoising. Note that it fails when we add more noises in Case 6.

\begin{table}[h]
	\centering
	\begin{tabular}{llll}
		\hline
		{\bf }  &{\bf Case 4} & {\bf Case 5} &{\bf Case 6} \\
		\hline
		Before  &  0.372(0.422)  &   0.152(0.301)  &  0.0567(0.139)  \\
		After   &  0.690(0.441)  &  0.385(0.454)  &  0.224(0.365) \\
		\hline
	\end{tabular}
	\caption{Adjusted Rand Index for Spiral Datasets.}
	\label{tab::sp}
\end{table}

We summarized the results of ARI for Case 4-6 in Table \ref{tab::sp}. 
In all cases, we see a clear improvement in the clustering performace even when the noise ratio is high.

\begin{table*}[h]
	\centering
	{\begin{tabular}{ccccc}
			\hline
			{\bf Dataset}  &{\bf MSD} &{\bf K-means} & {\bf Spectral clustering} &{\bf Hierarchical clustering} \\
			\hline
			Olive Oil & Before  &  0.635  &   0.621  &  0.815  \\
			& After   &  0.807  &  0.707  &  0.837 \\
			\hline
			Bank Authentication & Before  &  0.210  &   0.637  &  0.062  \\
			& After   & 0.233  &  0.708  &  0.108 \\
			\hline
			Seeds & Before  &  0.773  &   0.470  &  0.686 \\
			& After   & 0.798  &  0.742  &  0.585 \\
			\hline
		\end{tabular}}
		\caption{Adjusted Rand Index for Three Real Datasets. Note that the selection of bandwidth and the number of clusters are based on the previous work in \citet{chen2016comprehensive}. The spectral clustering gives a random result each time due to its implicite use of k-means clustering and here we just display one output.}
		\label{tab::realdata}
	\end{table*}
	
\subsubsection{Real Data Analysis}

In this subsection, we demonstrate the improvement brought by the MSD 
using three real datasets: the olive oil data, the banknote authentication data, and the seed data.
Note that the bandwidth $h$ and the number of groups $k$ are chosen by the results in \citet{chen2016comprehensive}.

\textbf{Olive Oil data.} This dataset is introduced in \citet{forina1983classification}, which consists of $d=8$ features and $n=572$ observations. The chosen bandwidth is $h=0.587$ and the number of groups is $k=7$.

\textbf{Banknote Authentication Data.} The data is from the UCI machine learning database repository \Citep{asuncion2007uci}. It contains $n=1372$ observations, each observation has $d=4$ attributes. The bandwidth and number of groups are respectively $h=0.453$ and $k=5$.

\textbf{Seeds Data.} The data is also from the UCI machine learning database repository \Citep{asuncion2007uci}, which includes $n=210$ observations, each has $d=7$ attributes. The bandwidth is $h=0.613$ and the number of groups is $k=3$.


The clustering algorithms we apply include k-means clustering, spectral clustering and hierarchical clustering.
We compare the ARI before and after denoising and show the results in Table \ref{tab::realdata}.
In general, the clustering performance has been improved after applying the MSD. 
The only exception is in Seeds data; hierarchical clustering becomes even worse after the denoising
but the spectral clustering is greatly improved in this case.

\subsection{Enhancements in Two-sample Tests} 
Now we show how the MSD improves the power of two-sample tests.  
The goal of two-sample tests is to determine whether two given samples are from the same distribution. Here we consider the Energy test \citep{rizzo2008energy,szekely2004testing,szekely2013energy} 
and the Kernel test \citep{gretton2012kernel} for illustration. 
Again we apply the empirical mean shift only one time to denoise the datasets.
The datasets are generated from Gaussian mixture model with two components, i.e.
\begin{equation}
p(x)=\pi \phi(x,\mu_1,\sigma_1^2) + (1-\pi) \phi(x,\mu_2,\sigma_2^2).
\label{eq::2samptest}
\end{equation}
The power is calculated by independently repeating data generation and two-sample tests 200 times. We consider two scenarios below.

\begin{figure*}[h]
	\centering
	\begin{subfigure}[b]{0.49\textwidth}
		\centering
		\includegraphics[width=5.5cm]{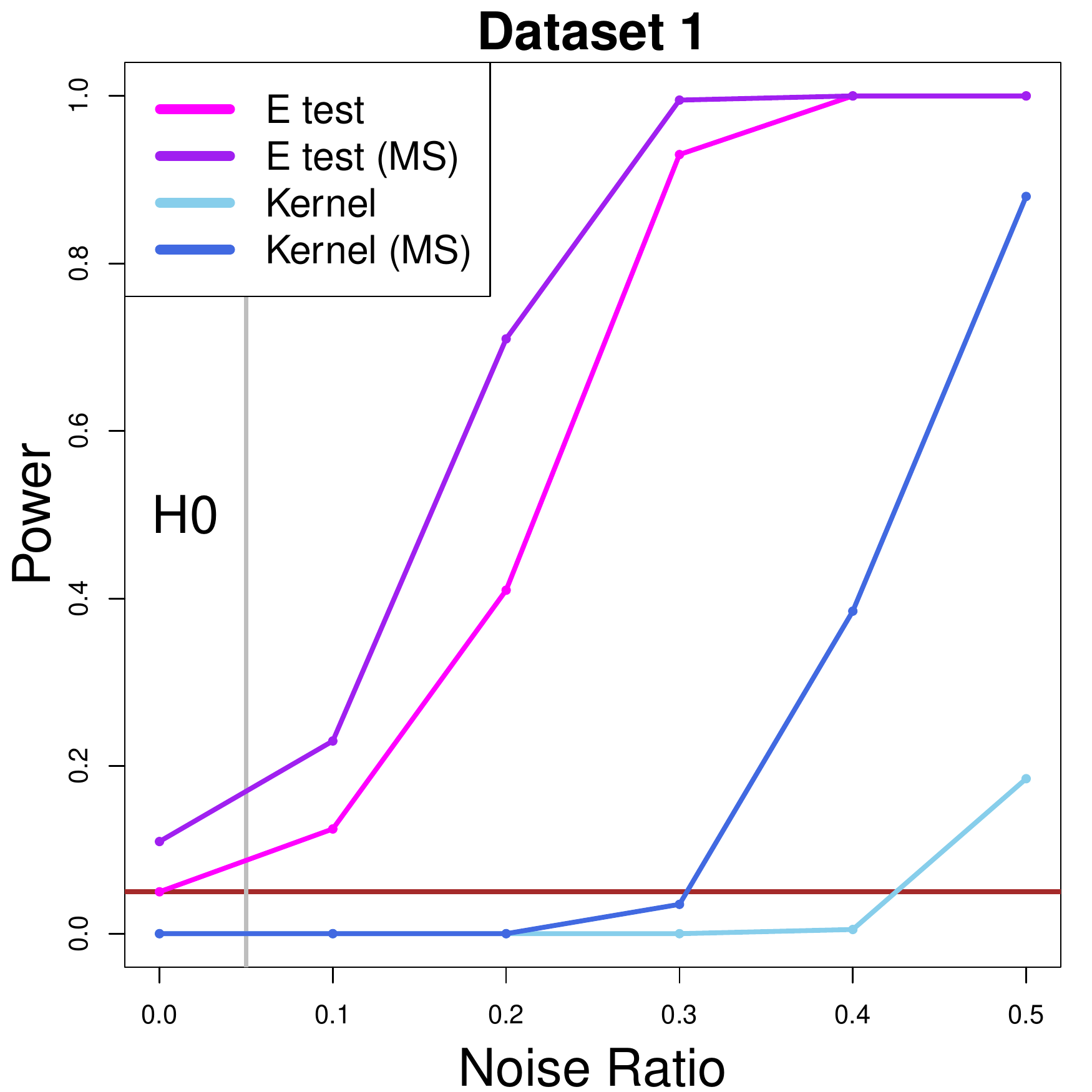}
		\caption{Uniform Noise}
		\label{fig::2samptest1}
	\end{subfigure}
	\begin{subfigure}[b]{0.49\textwidth}
		\centering
		\includegraphics[width=5.5cm]{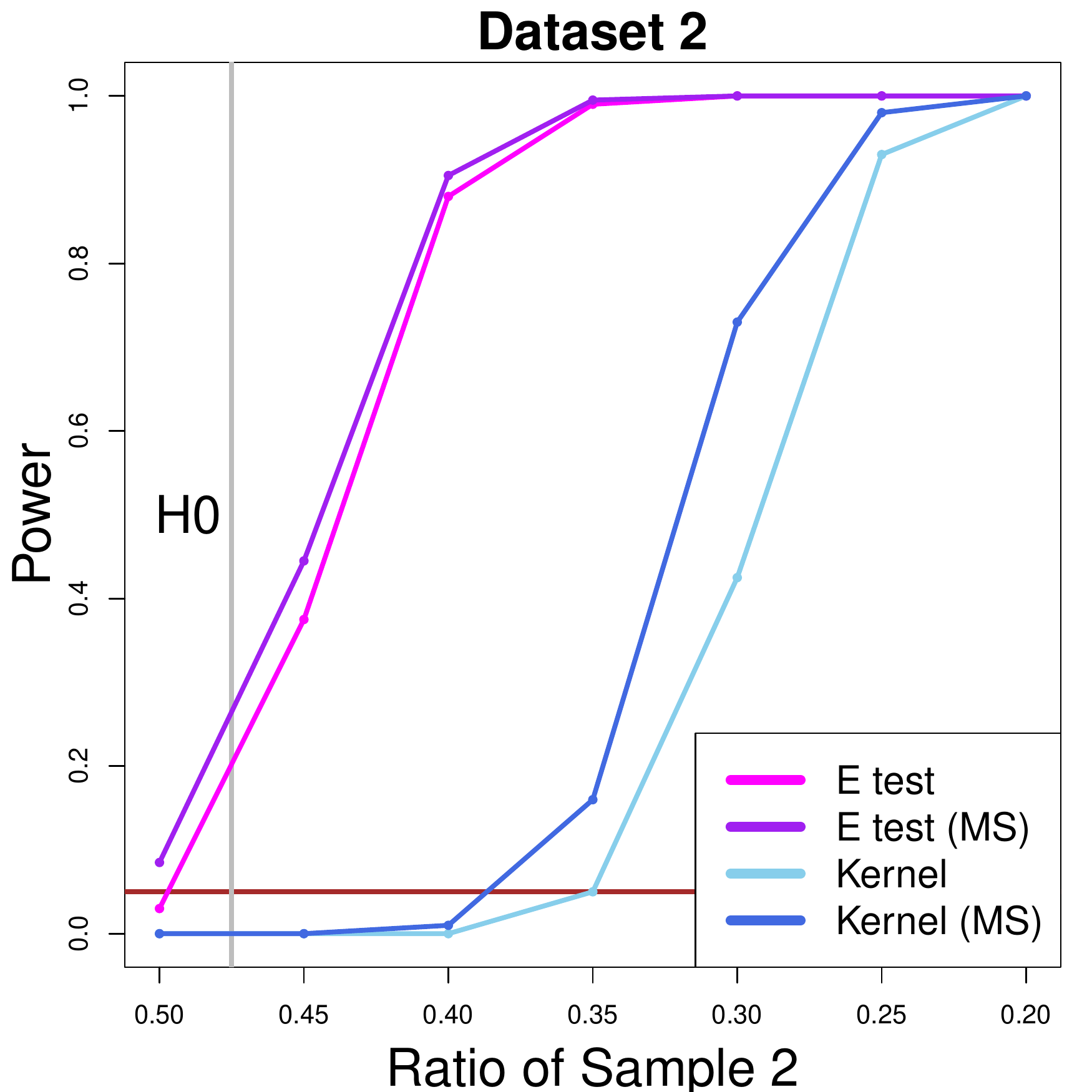}
		\caption{Different Mixture Proportion}
		\label{fig::2samptest3}
	\end{subfigure}
	\caption{Power curves of two-sample tests. In Dataset 1, we add extra uniform noises into S2 with increasing size. In Dataset 2, we change the mixture proportion of two components in S1. Area at the left side of the gray line represents $H_0$: two samples are from the same distribution. Area at the right side of the line is for $H_1$: two samples are from different distributions. The red line is for power=0.05.}
	\label{fig::2samptest}
\end{figure*}

\textbf{Uniform Noise.}  In this case, we show how the power changes with increasing noise before and after denoising. The datasets include two samples, S1 and S2, which are generated according to \eqref{eq::2samptest} and set the parameters as
$$
\pi=0.7, ~~\mu_1=0, ~~\mu_2=5, \sigma_1=\sigma_2=1.
$$
Both samples have $N_0=1000$ data points. We include additional uniform noise to S2 with size $N_1$ increasing from 0 to 500 by 100. Obviously, as more uniform noise is added, the greater differences these two samples have. The results are displayed in Figure \ref{fig::2samptest}.

\textbf{Various Mixture Proportion.}  In this case, we decrease the $\pi$ of S1 from 0.5 to 0.2 by 0.05 while keeping $\pi=0.5$ constant in S2 to see whether the MSD can help to distinguish two samples with increasing difference between them. See Figure \ref{fig::2samptest3}.

In Figure \ref{fig::2samptest}, we plot the power curve under $H_0$ and $H_1$. In general, Energy test has a better performance than Kernel MMD test as powers under $H_1$ are much higher, i.e. Energy test is more sensitive to the tiny discrepancies in two samples and thus has a lower false-positive rate. However, due to its sensitivity, the power of Energy test after MSD is larger than 0.05 under $H_0$. 
This is because the KDEs of the two samples are different due to the randomness of the sample so that 
the empirical mean shift moves the two sample toward slightly different targets, increasing the type 1 error (power under $H_0$) rate;
note that when sample size increases, the two KDEs will converge to the same target under $H_0$ so we would not have this issue.

\subsection{Anomaly Detection}
\begin{figure}
	\centering
	\includegraphics[scale=0.35]{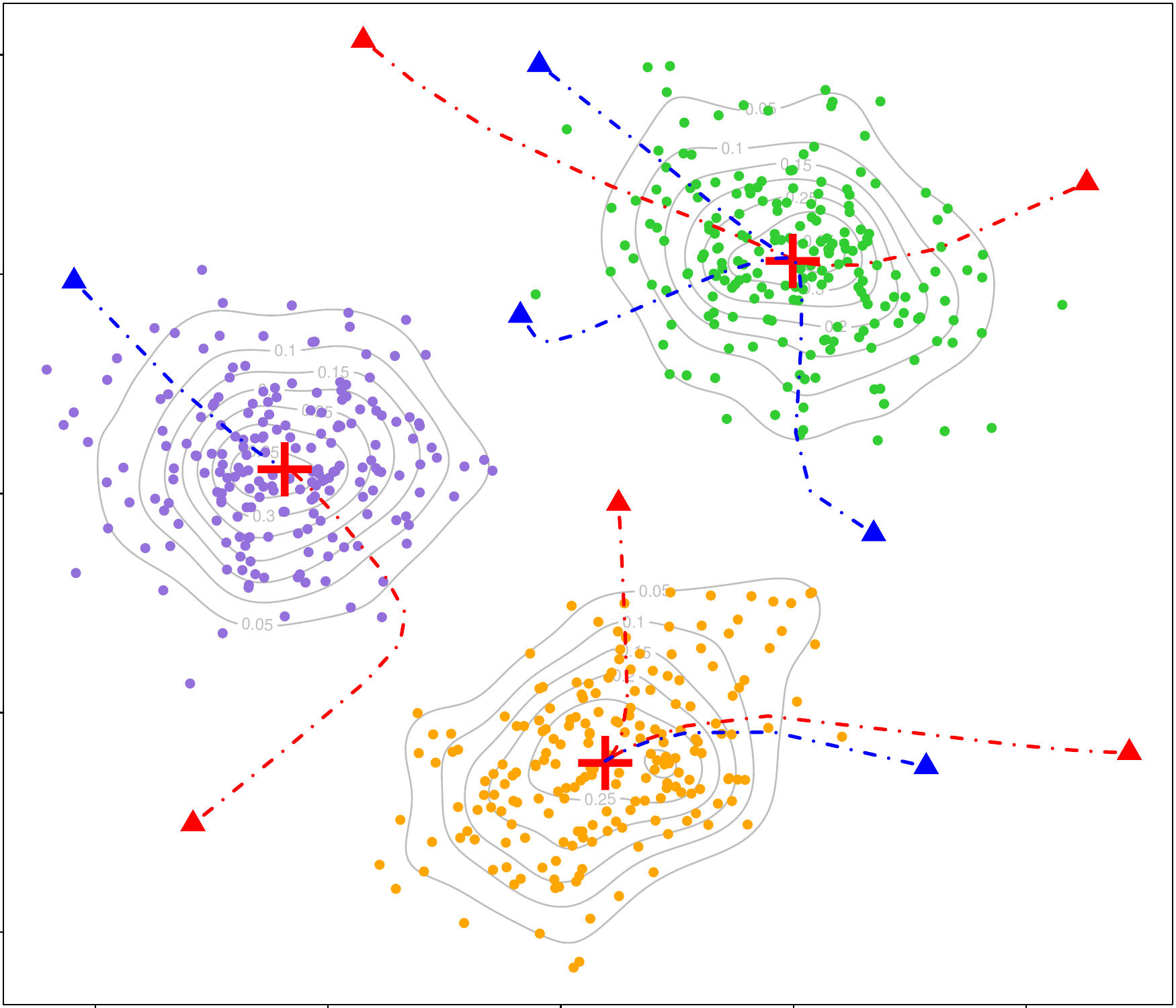}
	\caption{Anomaly detection. Three piles of Gaussian points with five artificial outliers. The red bold cross-overs represent the local mode estimated by kernel density. The red triangles are the five artificial outliers. The blue triangles are the other five points located in the low density regions and discovered by our anomaly detection approach.}
	\label{fig::outliers}
\end{figure}

Finally, we show that the
MSD can also be used to detect anomaly points (outliers). 
Here we propose a simple method for anomaly detection. 
For each data point, we perform the mean shift algorithm until it converges and
use the total shifted length as an anomaly index.
We demonstrate this idea using the example in Figure \ref{fig::outliers}
where
we generate a dataset using a Gaussian mixture model with $3$ components and
each component contains $200$ data points.
Then, we artificially put $5$ outliers within the low density area (marked as the red triangles). 
We then iterate the meanshift algorithm until convergence and record the total shifted length for every point.  
We find out the top $10$ points with the highest anomaly index and plot their traces in Figure \ref{fig::outliers}. 
The red triangles and traces are the actual outliers we added in; in this case we do successfully recover all of them.
The blue triangles are traces are the identified anomaly points that are from the Gaussian mixture model;
despite these points are from the Gaussian mixture model, they are also in the low density area
so classifying them as anomaly points is reasonable.


\section{Discussion}	\label{sec::discussion}
In this paper, we propose to use the mean shift algorithm as a denoising procedure.
We introduce a framework for analyzing how the mean shift changes a distribution
and show that the concentration rate of density at local modes and the probability mass within a level set increases by order $O(h^2)$
when we shift the data points once. 
We then apply the idea of the MSD to clustering, two-sample test, and anomaly detection to show that
all these statistical analysis can be improved by the MSD. 

There are many possible future extensions based on this paper.
For instance, the subspace constraint mean shift algorithm \citep{ozertem2011locally}
is a modified method of the usual mean shift algorithm that moves
data points toward ridgelines of the density function \citep{chen2015asymptotic,Genovese2012a}. 
This approach can be used to denoise, and both theoretical performance and applications in data analysis
can also be analyzed via a similar framework as this paper.

\bibliographystyle{plainnat}

\bibliography{paper.bib}

\appendix

\section{Proof of Theorem~\ref{thm::g_mode}}

Theorem \ref{thm::QP} and \ref{thm::mode} are special cases of Theorem~\ref{thm::g_mode}, so we only need to prove Theorem~\ref{thm::g_mode}. 
Note that in Theorem~\ref{thm::QP},  we pick $\rho_0 = \frac{1}{2}\lambda$ and $\epsilon_0 = \frac{\lambda}{3g_0}$. 

Since Theorem~\ref{thm::g_mode} has three conclusions: local modes, local minima, and level sets. We separately prove each part. 
We first introduce the concept of the geometric density \cite{mattila1999geometry,preiss1987geometry}.

\begin{definition} (Geometric Density) 
	Let $X\in \mathbb{R}^d$ be a random variable from a probability distribution $P$, then the geometric density function $\rho(x)$ is defined as
	$$
	\rho(x) = \lim_{r\rightarrow 0}\frac{P(B(x,r))}{v_d \cdot r^d},
	$$
	where $B(x,r)$ is a closed ball of radius $r$ centered at $x$, $v_d$ is a constant and represents the volume of an unit ball in $d$-dimensional space.
	\label{def::geoden}
\end{definition}
Note that when the usual density (also called the Lebesgue density) is finite, the geometric density equals to the usual density \citep{mattila1999geometry}.

Recall that $S$ is the probability distribution of the shifted variable $X'$ by $\mathbb{M}_{f,\tau}$, see \eqref{eq::g_shift_op} and $s$ is the associated density function. 
It is easy to see that $s$ is finite so the usual density equals to the geometric density.
Then based on Definition \ref{def::geoden}, the density at $x$ after denoising is
\begin{equation}
s(x) = \lim_{r\rightarrow 0}\frac{S(B(x,r))}{v_d \cdot r^d}.
\label{eq::def::s}
\end{equation}
Thus, to see how $s(x)$ is different from $p(x)$, we need to investigate how $S(B(x,r))$ changes when $r\rightarrow 0$.

\subsection{Density at Local Modes}

We first consider the case of local modes.
Let $m$ be a local mode of $f$
and $A_r$ be a collection of points that will be shifted into $B(m,r)$, i.e.
\begin{equation}
	A_r = \left\{x: M_{f,\tau}(x) = x+c\cdot \tau^2\cdot\frac{\nabla f(x)}{f(x)}\in B(m,r)\right\}, 
	\label{eq::Ar}
\end{equation}
	
By the definition of $S$ and $A_r$, it is easy to see
	\begin{equation}
	S(B(m,r)) = P(A_r).
	\label{eq::Ar2}
	\end{equation}
Therefore, all we need is to understand the rate of $P(A_r)$ as a function of $r$. 
Let $A_r/B=\{x: x\notin B(m,r) \text{ and } x\in A_r\}$, then $P(A_r)$ can be bounded by
\begin{equation}
\begin{aligned}
	P(B(m,r_{\min})) \le &P(A_r) \le P(B(m,r_{\max})),\\
	\Rightarrow B(m,r_{\min}) \subset &A_r \subset B(m,r_{\max}).
\end{aligned}
\label{eq::bound}
\end{equation}
where $r_{\max} = \sup_{x\in A_r/B}\{\|x-m\|\}$ and $r_{\min} = \inf_{x\in A_r/B}\{\|x-m\|\}$. 
Thus, we need to find $r_{\max}$ and $r_{\min}$.
	
To make sure $x\in A_r/B$, the shifted distance must be more than the distance from $x$ to $B(m,r)$, i.e. the following inequality must always hold,
\begin{equation}
	\left\|c\cdot \tau^2 \cdot\frac{\nabla f(x)}{f(x)}\right\| =c\cdot \tau^2 \cdot\frac{\left\|\nabla f(x)\right\|}{f(x)}  \geq \|x-m\|-r>0.
	\label{eq::mode::shift}
\end{equation}
We will use this equation in deriving the upper bound and the lower bound on the increase of density.

\textbf{Upper bound.} 
When $r\rightarrow 0$,
$A_r$ shrinks to a tiny region around $m$.
In this case, for any point $x\in A_r$, by assumption (A1) the density and the gradient at $x$ could be approximated by 
\begin{equation}
f(x) = f(m)+O(\|x-m\|^2)
\label{eq::approx1}
\end{equation}
\begin{equation}
\|\nabla f(x)\| \approx \|\nabla\nabla f(x)(x-m)\|\leq \|f\|_{2,\max} \|x-m\|
\label{eq::approx2}
\end{equation}

To obtain a upper bound for $\|x-m\|$, we need increase the left hand side (LHS) in \eqref{eq::mode::shift} as much as possible. Thus, with the approximation in \eqref{eq::approx1} and \eqref{eq::approx2},  we have
\begin{equation}
	\|x-m\|-r \leq c\cdot \tau^2 \cdot\frac{\|f\|_{2,\max} \|x-m\|}{f(m)}.
	\label{eq::mode::res}
\end{equation}
After rearrangement, 
	$$
	\|x-m\| \leq r/\left(1-\frac{c\cdot \tau^2\cdot \|f\|_{2,\max}}{f(m)}\right) = \frac{r}{1-\tau^2 \cdot c^*},
	$$
where $c^* = \frac{c\cdot \|f\|_{2,\max}}{f(m)}$.
	Thus, we choose $r_{\max}=\frac{r}{1-\tau^2 \cdot c^*}$ which leads to 
	$$
	A_r \subset B\left(m, \frac{r}{1-\tau^2 \cdot c^*}\right) = B\left(m, r_{\max}\right)
	$$
By equation \eqref{eq::Ar2} and \eqref{eq::def::s},
\begin{align*}
	s(m) &= \lim_{r\rightarrow 0}\frac{S(B(m,r))}{v_d \cdot r^d}\\
	& = \lim_{r\rightarrow 0}\frac{P(A_r)}{v_d \cdot r^d}\\
	&\leq \lim_{r\rightarrow 0}\frac{P\left(B\left(m,\frac{r}{1-\tau^2\cdot c^*}\right)\right)}{v_d \cdot r^d}\\
	& = \lim_{r\rightarrow 0}p(m)(1+O(r))\left(\frac{1}{1-\tau^2\cdot c^*}\right)^d\\
	& = \frac{p(m)}{(1-\tau^2\cdot c^*)^d}\\
	&\leq p(m)(1+\tau^2\cdot c^{\dagger}),
\end{align*}
where $c^{\dagger}$ is some constant depends only on $d, c^*$ when $\tau$ is sufficiently small. 

This proves the upper bound for the case of local modes. Note that equation \eqref{eq::mode::res} is valid when $c\cdot\tau^2\cdot \|f\|_{2,\max}<f(m)$, 
which gives us a restriction on $\tau:\,\,\tau^2< \frac{f(m)}{c\cdot \|f\|_{2,\max}}$.

\textbf{Lower bound.} The derivation of the lower bound uses a similar idea as the upper bound. But now we consider the $x$ on the direction of the eigenvector corresponds to the smallest absolute eigenvalue $\lambda_{\min}$.
Note that $\lambda_{\min}>0$ because of assumption (A2).

Let $x \in A_r$. To get the lower bound, we choose $x$ such that
$x-m$ is parallel to the eigenvector of $\nabla \nabla f(x)$ corresponding to eigenvalue $\lambda_{\min}$. 
When $r$ is small, the amount of gradient $\|\nabla f(y)\| \geq \frac{\lambda_{\min}}{2}\|y-m\|$ for every $y\in A_r$. Note that $\frac{\lambda_{\min}}{2}$ is half of the smallest eigenvalue; we need a factor $2$ because when $x$ is away from $m$, the second derivative of $f(x)$ may change the eigenvalues. 

Thus, for point $x$, when the shifted distance is more than $\|x-m\|-r$, it will be shifted into the ball $B(m,r)$. By equation \eqref{eq::mode::shift}, the shifted distance
\begin{equation}
	c\cdot \tau^2 \cdot\frac{\left\|\nabla f(x)\right\|}{f(x)} \geq c\cdot \tau^2 \cdot\frac{\lambda_{\min}\|x-m\|}{2f(m)};
	\label{eq::lower::shift}
\end{equation}
If we want a lower bound of $\|x-m\|$, we need the LHS of \eqref{eq::mode::shift} as small as possible. 
Using the same approximation in \eqref{eq::approx1}-\eqref{eq::approx2} and equation \eqref{eq::lower::shift}, we can know that 
$x$ will be shifted into $B(m,r)$ when the following condition holds
\begin{equation}
	c\cdot \tau^2 \cdot\frac{\lambda_{\min}\|x-m\|}{2f(m)} \geq \|x-m\| - r.
	\label{eq::mode::res2}
\end{equation}
After rearrangement, the above equation equals to
$$
\|x-m\| \leq r/\left(1-\frac{c\cdot \tau^2\cdot \lambda_{\min}}{2f(m)}\right)
$$
This gives us a lower bound $r_{\min}=\frac{r}{1-\tau^2\cdot c_{*}}$, where $c_{*}=\frac{c\lambda_{\min}}{2f(m)}$. Thus
$$
B\left(x, r_{\min}\right)=B\left(x, \frac{r}{1-\tau^2\cdot c_{*}}\right)\subset A_r.
$$

Now by the same derivation as the upper bound, we obtain a lower bound on the density $s(m) \geq p(m)(1+\tau^2\cdot c_{\dagger})$, where $c_{\dagger}$ is some constant depending only on $d$ and $f$ when $\tau$ is sufficiently small. This proves the result of lower bound. 

Note that inequality \eqref{eq::mode::res2} hold under $\tau^2< \frac{2f(m)}{c\cdot \lambda_{\min}}$. This gives another sets of restriction on $\tau$  (and this restriction of $\tau$ is tighter than the one from the upper bound).

\subsection{Density at Local Minima}

The proof for the case of local minima follows a similar derivation as the case of local modes so we ignore the proof. 
The only difference is that the set $A_r =\left\{x: x+c\cdot \tau^2\cdot\frac{\nabla f(x)}{f(x)}\in B(m,r)\right\} \subset B(m,r)$, where $m$ represents a local minimum.

\subsection{Probability Mass in Level Sets}
	
Given a level set $L_\lambda = \{x: f(x)\geq \lambda\}$, to investigate how the shifted distribution concentrates, we need to to study the following region:
$$
B_\lambda = \left\{x: x+c\cdot \tau^2 \cdot \frac{\nabla f(x)}{f(x)} \in L_\lambda\right\}.
$$
Namely, $B_\lambda$ is the collection of regions that will be shifted into $L_\lambda$ after applying the generalized mean shift algorithm once.
	
Now consider a point $x$ close to $L_\lambda$ but $x\notin L_\lambda$. 
Let $\delta(x) = d(x,L_\lambda)$ be the minimal distance from $x$ to the level set $L_\lambda$ and let $\pi(x) \in \partial L_\lambda$ be the projected point from $x$ to $L_\lambda$. Note that $\delta(x) = \|x-\pi(x)\|$.

\begin{figure}
	\center
	\includegraphics[width=2in]{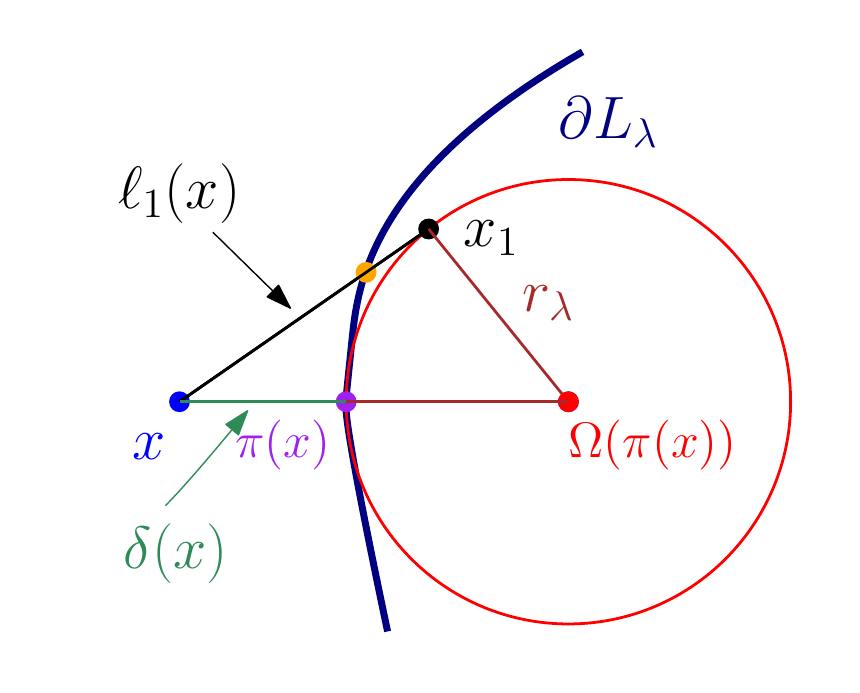}
	\caption{A diagram to illustrate the proof of the case of level sets.
		The blue dot is the point $x$, red dot is $\Omega(\pi(x))$, the center of the ball intersecting 
		the boundary $\partial L_\lambda$ at the projected point $\pi(x)$.
		Thus, the length of the green line segment is $\delta(x)$,
		the length of the two brown line segments are $r_\lambda$, the reach of $L_\lambda$.
		$x_1$ is the first intersecting point on the ball from the line from $x$ moving toward $\nabla f(x)$
		and the black line segment is $\overline{xx_1}$, whose length is $\ell_1(x)$.
		Note that $\ell_0(x)$ is the length of the line segment between $x$ and the orange point on $\overline{xx_1}$. 
		The angle $\theta$ is the angle between $\overline{xx_1}$ and $\overline{x\pi(x)}$.
	}
	\label{fig::pf::lv1}
\end{figure}

The main idea of the proof is to show that when $\delta(x)\leq c_0 \cdot \tau^2$ for some fixed constant $c_0$, $x\in B_\lambda$. A key observation is that this occurs when the shifted distance $c\cdot \tau^2 \cdot \frac{\|\nabla f(x)\|}{f(x)}$ is greater than $c_1 \cdot \delta(x)$ for some constant $c_1\geq1$. Namely, as long as $x$ is shifted for long enough distance, the shifted position is inside $L_\lambda$. The reason why the constant $c_1\geq1$ is because the shift orientation is along $\nabla f(x)$, which might not be the same as $\pi(x)-x$ (the direction of shortest path) so we need to take this into account.
	
To derive the constant $c_0$, we first introduce a useful quantity called `reach' \cite{Federer1959}. 
\begin{definition}
	Given a set $A$, the reach of $A$ is the largest distance from $A$ such that every point within this distance to $A$ has a unique projection onto $A$. That is
	$$
	 \text{reach}(A)=\sup\{r: \pi_A(x) \text{ is unique } \forall x \in A \oplus \epsilon\}, 
	$$
	where $A\oplus \epsilon = \bigcup_{x\in A}B(x,\epsilon)$.
	\label{def::reach}
\end{definition}
Intuitively, if a set $A$ has ${\sf reach}(A)=r_0$, then we can put a ball with radius $r_0$ and roll it freely on the boundary of set $A$. 
This also implies that we can freely move a ball with radius $r_0$ within the set $A$ without penetrating $A$.
We refer the readers to \cite{Federer1959, chen2015density} for more discussion about reach.
	
Let $r_\lambda$ be the reach of $L_\lambda$.  
For any point $y\in \partial L_\lambda$, there is a ball $B(\Omega(y), r_\lambda)\subset L_\lambda$ such that $y\in B(\Omega(y),r_\lambda)$ and $\Omega(y) = y + r_\lambda \cdot \frac{\nabla f(y)}{\|\nabla f(y)\|}$. Namely, the ball $B(\Omega(y), r_\lambda)$ intersect the boundary of level set $\partial L_\lambda$ at one and only one point $y$. Note that $\frac{\nabla f(y)}{\|\nabla f(y)\|}$ is the normal vector to $\partial L_\lambda$ at point $y$.
From Lemma 1 in \cite{chen2015density} and Assumption (A3), the reach of $L_\lambda$, $r_\lambda\geq \frac{g_0}{\|f\|_{2,\infty}}$.
	
For the point $x$, it will be shifted along the direction $\nabla f(x)$. Let $\theta$ be the angle between two vectors $\nabla f(x)$ and $\pi(x)-x$. We assume $\theta$ is very small (later we will derive an upper bound for $\theta$); this occurs when $\delta(x)$ is small. Because $\pi(x)-x$ is in the same direction as $\nabla f(\pi(x))$ (both are in the direction of $\pi (x) - m(\pi(x))$), then we have
$$
\nabla f(x)  =  \nabla f(\pi(x))  + \int_{z=\pi(x)}^{z=x} \nabla \nabla f(z),
$$
from which we obtain $\|\nabla f(x)-\nabla f(\pi(x))\|\leq \|f\|_{2,\infty}\|x-\pi(x)\| =  \|f\|_{2,\infty}\delta(x)$. Thus, the largest $\theta$ is achieved when the differences are all in the perpendicular direction,
	which implies
	$$
	\tan(\theta) \leq \frac{\|f\|_{2,\infty} \delta(x)}{\|\nabla f(x)\|} = \tan(\overline{\theta}),
	$$
	where $\overline{\theta}$ is the upper bound for $\theta$.
	
Now let $B(\Omega(\pi(x)), r_\lambda)\in L_\lambda$ be the ball intersecting $\partial L_\lambda$ at point $\pi(x)$, where $\Omega(\pi(x)) = \pi(x) + r_\lambda \cdot \frac{\nabla f(\pi(x))}{\|\nabla f(\pi(x))\|}$.  Let $\ell_0(x)$ be the distance such that $x+ \ell_0(x) \cdot \frac{\nabla f(x)}{\|\nabla f(x)\|} \in L_\lambda$. That is, $\ell_0(x)$ is the distance we need to shift so that $x\in L_\lambda$. Let $\ell_1(x)$ be the distance from $x$ along direction $\nabla f(x)$ to intersect $B(\Omega(\pi(x)),r_\lambda)$;
let the first intersecting point be $x_1$. Then it is easy to see $\ell_0(x)\leq \ell_1(x)$. Meanwhile,  $x, \Omega(\pi(x))$, and $x_1$ form a triangle with $\theta$ being the angle of $\overline{x\Omega(\pi(x))}$ and $\overline{xx_1}$ where 
$$\overline{x\Omega(\pi(x))} = \delta(x)+r_\lambda, ~~\overline{xx_1}=\ell_1(x),  ~~\|\overline{\Omega(\pi(x))x_1}\|=r_\lambda,$$
see Figure \ref{fig::pf::lv1} for illustration. Therefore, based on the law of cosine, 
\begin{equation}
	(\delta(x) +r_\lambda)^2 + \ell_1(x)^2 - 2\ell_1(x)\cdot(\delta(x) +r_\lambda)\cdot \cos(\theta) = r_\lambda^2.
	\label{eq::law_cosine}
\end{equation}
By solving equation \eqref{eq::law_cosine}, the expression of $\ell_1(x)$ is
\begin{equation}
	\ell_1(x)  = A -\sqrt{A^2-B}
	\label{eq::ell1_1}
\end{equation}
where $\quad A = (\delta(x)+r_\lambda)\cos(\theta),\quad B = \delta^2(x)+2\delta(x)r_\lambda$. 
Note that the other solution does not make sense since $\theta$ is small.


To derive an explicit bound, we assume
	\begin{equation}
	\cos^2(\theta)\geq\frac{1}{2} ,\quad r_\lambda \geq \delta(x);
	\label{eq::bound_r}
	\end{equation}
and later we will show that this occurs when $\tau^2$ is sufficiently small.
	Using the fact that $\sqrt{1-x}>1-x$ when $0<x<1$ and equation \eqref{eq::bound_r} and \eqref{eq::ell1_1}, 
	\begin{equation}
	\begin{aligned}
	\ell_1(x) &= A-\sqrt{A^2-B}\\
	& = A - A\sqrt{1-\frac{B}{A^2}} \\
	&\leq A - A(1-\frac{B}{A^2})\\
	& = \frac{B}{A}\\
	& = \frac{\delta^2(x)+2\delta(x)r_\lambda}{(\delta(x)+r_\lambda)\cos(\theta)}\\
	& \leq \sqrt{2}\delta(x) + \frac{\sqrt{2}\delta(x)}{\delta(x)/r_\lambda + 1}
	\end{aligned}
	\end{equation}
		
	
The inequality always holds when 
\begin{equation}
\begin{aligned}
\ell_1(x) 
&\leq \inf_{r_\lambda \geq \delta(x)}\{\sqrt{2}\delta(x) + \frac{\sqrt{2}\delta(x)}{\delta(x)/r_\lambda + 1}\}\\
& = \frac{3\sqrt{2}\delta(x)}{2}
\end{aligned}
\end{equation}

Because $\ell_0(x)\leq \ell_1(x)$, this implies an upper bound $\ell_0(x) \leq 3\sqrt{2}\delta(x)/2$, which shows that the constant $c_1 = 3\sqrt{2}/2$. 
	
Thus, when the shifted distance is more than $3\sqrt{2}\delta(x)/2$, $x\in B_\lambda$. Namely,
$$
\frac{3}{2}\sqrt{2}\delta(x)<c\cdot \tau^2 \cdot \frac{\|\nabla f(x)\|}{f(x)} \Longrightarrow x\in B_\lambda.
$$
For $x\notin L_\lambda$ and $\delta(x)\leq\frac{g_0}{2\|f\|_{2,\infty}}$,  the above inequality holds whenever
$$
\frac{3}{2}\sqrt{2}\delta(x)<c\cdot \tau^2 \frac{g_0}{2\lambda}.
$$
This is from the fact that $\nabla f(x) >g_0$ and $f(x) < \lambda$ for $x\notin L_\lambda$. Thus, the set 
	\begin{equation}
	C_\lambda(\tau) = \left\{x: x\notin L_\lambda, \delta(x) < c\cdot \tau^2 \frac{g_0}{3\sqrt{2}\lambda}\right\} \subset B_\lambda.
	\label{eq::C}
	\end{equation}
	Because $C_\lambda \cap L_\lambda=\phi$, we have
	$$
	Q(L_\lambda) - P(L_\lambda) = P(B_\lambda) - P(L_\lambda) \geq P(C_\lambda \cup L_\lambda) - P(L_\lambda) = P(C_\lambda).
	$$
If $\delta(x)< \epsilon_0$, then $p(x)\geq \rho_0$ for all $x\in C_\lambda$.
	Thus,
	\begin{align*}
	P(C_\lambda) &= \int_{x\in C_\lambda} p(x) dx\geq p_0 {\sf Vol}(C_\lambda)\\ 
	&\geq \rho_0 \cdot {\sf Vol}_{d-1}(\partial C_\lambda) \cdot c\cdot \tau^2 \frac{g_0}{3\sqrt{2}\lambda}\\
	&\geq \rho_0 \cdot {\sf Vol}_{d-1}(\partial L_\lambda) \cdot c\cdot \tau^2 \frac{g_0}{3\sqrt{2}\lambda}
	\end{align*}
which proves the probability bound.
Note that the last inequality follows from the fact that $L_\lambda \subset C_\lambda$ and $L_\lambda$ has reach at least $\frac{g_0}{2\|f\|_{2,\infty}}$
and $\inf_{x\in C_\lambda}d(x, L_\lambda) < \frac{g_0}{2\|f\|_{2,\infty}}$ so the set $C_\lambda $ is just an extended set of $L_\lambda$. 
Thus, $C_\lambda$ has a larger boundary than $L_\lambda$, which implies $\partial C_\lambda \geq \partial L_\lambda$.

To obtain the above bound ,we need equation \eqref{eq::C} and $\delta(x)< \epsilon_0$. To ensure equation \eqref{eq::C}, we assumed $\delta(x)\leq\frac{g_0}{2\|f\|_{2,\infty}}$ and equation \eqref{eq::bound_r}. Sufficient conditions of these assumptions are the following three inequalities:
	\begin{equation}
	\begin{aligned}
	\tan(\overline{\theta}) = \frac{\|f\|_{2,\infty} \delta(x)}{\|\nabla f(x)\|}&\leq 1\\
	\delta(x)&\leq  \frac{g_0}{\|f\|_{2,\infty}}\\
	\delta(x)&\leq\frac{g_0}{2\|f\|_{2,\infty}}
	\end{aligned},
	\label{eq::LV::h0}
	\end{equation}
which actually only requires $\delta(x)\leq  \frac{g_0}{2\|f\|_{2,\infty}}$ (note that $\|\nabla f(x)\|\geq \frac{1}{2}g_0$ whenever $\delta(x)\leq \frac{g_0}{2\|f\|_{2,\infty}}$). Because equation \eqref{eq::C} has assigned an upper bound of $\delta(x)$ using $\tau^2$, assumptions \eqref{eq::LV::h0} will be true if
	\begin{equation}
	\begin{aligned}
	&c\cdot \tau^2 \frac{g_0}{3\sqrt{2}\lambda} \leq \frac{g_0}{\|f\|_{2,\infty}}\\
	\Longleftrightarrow\quad& \tau^2 \leq \frac{3\sqrt{2}\lambda}{c\cdot\|f\|_{2,\infty}},
	\end{aligned}
	\label{eq::LV::h1}
	\end{equation}
	which is an upper bound we request for $\tau^2$.
	The other upper bound comes from the fact that if 
	$c\cdot \tau^2 \frac{g_0}{3\sqrt{2}\lambda}\leq \epsilon_0$, 
	$\delta(x)\leq \epsilon_0$.
	Thus, a sufficient condition for $\tau$ is 
	$$
	\tau^2 \leq \min\left\{\frac{3\sqrt{2}\lambda}{c\cdot\|f\|_{2,\infty}}, \frac{3\sqrt{2}\lambda\epsilon_0}{c\cdot g_0}\right\}.
	$$

\section{Proof of Theorem \ref{thm::perturb}}

Because Theorem~\ref{thm::QQ} and \ref{thm::estimate}
are special cases of Theorem~\ref{thm::perturb}, we only need to prove Theorem~\ref{thm::perturb}.

\subsection{Situation 1: $\{f_n\}, ~\Delta_{1,n}\rightarrow 0$}

By the definition of $S_{f,\tau,P}$, for any set $W$
$$
S_{f,\tau,P}(W) = P\left(\left\{x: x+c\cdot\tau^2\cdot \frac{\nabla f(x)}{f(x)}\in W\right\}\right) = P(D(f)),
$$ 
where $D(f) =  \left\{x: x+c\cdot\tau^2\cdot \frac{\nabla f(x)}{f(x)}\in W\right\}$.
Similarly, we define
$D(f_n) = \left\{x: x+c\cdot\tau^2\cdot \frac{\nabla f_n(x)}{f_n(x)}\in W\right\}$
which leads to 
$$
S_{f_n,\tau,P}(W) = P(D(f_n)).
$$
Thus, all we need is to study the difference between $D(f_n)$ and $D(f)$.

For a point $x$ and a set $A$, recalled that $d(x,A)$ is the projection distance from $x$ to $A$.
A feature between $D(f)$ and $D(f_n)$ is that for any point $x\in D(f)$, due to the triangular inequalities, 
$d(x,D(f_n)) \leq O(\tau^2\cdot\Delta_{1,n})$,
where $\Delta_{1,n} = \max{\|f-f_n\|_{\ell,\infty}:\ell = 0,1}$,
and vice versa.

Now we define two set operations. 
For a set $A$ and a value $r>0$, $A\oplus r = \{x: d(x,A)\leq r\}$.
For two sets $A$ and $B$, $A\backslash B = \{x: x\in A, x\notin B\}$.

Thus, the above projection property 
implies that 
$$
D(f_n)\subset D(f)\oplus( A_0\cdot\tau^2\cdot \Delta_{1,n}),\quad D(f)\subset D(f_n)\oplus( A_0\cdot\tau^2\cdot \Delta_{1,n})
$$
for some constant $A_0>0$.
This further implies 
\begin{align*}
D(f_n)\backslash D(f) &\subset \left(D(f)\oplus( A_0\cdot\tau^2\cdot \Delta_{1,n})\right)\backslash D(f)\\
D(f)\backslash D(f_n) &\subset \left(D(f_n)\oplus( A_0\cdot\tau^2\cdot \Delta_{1,n})\right)\backslash D(f_n).
\end{align*}

A simple geometric observation is that 
for a set $A$ with non-zero surface area (i.e. ${\sf Vol}_{d-1}(A)>0$), 
the volume of $(A\oplus r)\backslash A$ is at rate $O(r)$ when $r$ is small. 
And it is easy to see that $D(f)$ and $D(f_n)$ both have non-zero surface area.

Therefore, using the notation $A\backslash B = \{x: x\in A, x\notin B\}$ for two sets $A$ and $B$,
\begin{align*}
|S_{f,\tau,P}(W)-S_{f_n,\tau,P}(W)|  &= |P(D(f))-P(D(f_n))|\\
& \leq P(D(f)\backslash D(f_n)) +P(D(f_n)\backslash D(f))\\
& \leq \|p\|_{0,\infty}( {\sf Vol}(D(f)\backslash D(f_n)) + {\sf Vol}(D(f_n)\backslash D(f)))\\
& = O (\tau^2\cdot \Delta_{1,n}),
\end{align*}
which proves the first case.

\subsection{Situation 2: $\{\tau_n\}, ~|\tau_n-\tau|\rightarrow 0$}

This follows the same derivation as the case of $\{f_n\}$ so we ignore the proof.

\subsection{Situation 3: $\{P_n\}$, $|P_n(B)-P(B)|\rightarrow 0$}

For any given set $A$, let
$$
D =  \left\{x: x+c\cdot\tau^2\cdot \frac{\nabla f(x)}{f(x)}\in A\right\}.
$$
Then $S_{f,\tau,P}(A) = P(D)$ and $S_{f,\tau,P_n}(A) = P_n(D)$.
Thus,
$$
|S_{f,\tau,P_n}(A)-S_{f,\tau,P}(A)| = |P_n(D) - P(D)|,
$$
which proves the result.

\end{document}